\documentclass[preprint,10pt]{aastex}

\newcommand{\kms}       {\mbox{km~s$^{-1}$}} 
\newcommand{\whz}       {\mbox{W~Hz$^{-1}$}}
\newcommand{\iras}      {\emph{IRAS}}
\newcommand{\atca}      {\emph{ATCA}}
\newcommand{\vla}       {\emph{VLA}}
\newcommand{\iraf}      {\emph{IRAF}}
\newcommand{\miriad}    {\emph{MIRIAD}}
\newcommand{\splot}     {\emph{splot}}
\newcommand{\onedspec}  {\emph{onedspec}}
\newcommand{\Lsun}      {\mbox{$L_{\odot}$}}
\newcommand{\Msun}      {\mbox{$M_{\odot}$}}
\newcommand{\nLnfir}    {\mbox{${\nu}L_\nu$(60~$\mu$m)}}

\newcommand{\Snfir}     {\mbox{$S_\nu$(60~$\mu$m)}}
\newcommand{\Snhun}     {\mbox{$S_\nu$(100~$\mu$m)}}
\newcommand{\Lnrad}     {\mbox{$L_\nu$(4.8~GHz)}}
\newcommand{\Snrad}     {\mbox{$S_\nu$(4.8~GHz)}}
\newcommand{\oiii}      {\mbox{[\ion{O}{3}]~$\lambda$~5007~\AA}}

\begin{document}

\title{Radio-Excess \iras\ Galaxies: PMN/FSC Sample Selection}

\author{Catherine L. Drake\altaffilmark{1,2}, Peter J. McGregor\altaffilmark{1}, Michael A. Dopita\altaffilmark{1}, \and W. J. M. van Breugel\altaffilmark{3}}

\altaffiltext{1}{Research School of Astronomy and Astrophysics, The Australian National University, Cotter Rd, Weston, ACT 2611 Australia}
\altaffiltext{2}{Australia Telescope National Facility, PO Box 76, Epping, NSW 1710 Australia}
\altaffiltext{3}{Institute of Geophysics and Planetary Physics, Lawrence Livermore National Laboratory, L-413, P.O. Box 808, Livermore, CA 94550}

\email{cdrake,peter,mad@mso.anu.edu.au,vanbreugel1@llnl.gov}

\shorttitle{Radio-Excess \iras\ Galaxies}
\shortauthors{Drake et al.}

\protect\singlespace

\begin{abstract}

A sample of 178 extragalactic objects is defined by correlating the 60
\micron\ IRAS FSC with the 5 GHz PMN catalog. Of these, 98 objects lie
above the radio/far-infrared relation for radio-quiet objects.  These
radio-excess galaxies and quasars have a uniform distribution of radio
excesses with no evidence for a radio-loud correlation equivalent to
the well known radio-quiet radio/far-infrared relation. The
radio-intermediate objects appear to be a new population of active
galaxies not present in previous radio/far-infrared samples chosen
using more stringent far-infrared criteria. The radio-excess objects
extend over the full range of far-infrared luminosities seen in
extragalactic objects, from low luminosity galaxies with $\nLnfir <
10^{9} \> \Lsun$ to ultra-luminous infrared galaxies with $\nLnfir >
10^{12}\> \Lsun$.  Objects with small radio excesses are more likely
to have far-infrared colors similar to starbursts, while objects with
large radio excesses have far-infrared colors typical of pure
AGN. Some of the most far-infrared luminous radio-excess objects have
the highest far-infrared optical depths.  These are good candidates to
search for hidden broad line regions in polarized light or via
near-infrared spectroscopy. Some low far-infrared luminosity
radio-excess objects appear to derive a dominant fraction of their
far-infrared emission from star formation, despite the dominance of
the AGN at radio wavelengths. Many of the radio-excess objects have
sizes likely to be smaller than the optical host, but show optically
thin radio emission, rather than flat radio spectra indicative of
compact quasar cores.  We draw parallels between these objects and
high radio luminosity Compact Steep-Spectrum (CSS) and GigaHertz
Peaked-Spectrum (GPS) objects.  Radio sources with these
characteristics may be young AGN in which the radio activity has begun
only recently. Alternatively, high central densities in the host
galaxies may be confining the radio sources to compact sizes.  We
discuss future observations required to distinguish between these
possibilities and determine the nature of radio-excess objects.

\end{abstract}


\section{Introduction}
\label{sec:int}

Active galaxies have traditionally been separated into radio-quiet and
radio-loud classes \citep{mil90}.  Radio-quiet objects include
starburst galaxies, Seyfert galaxies, and radio-quiet quasars.
Radio-loud objects include radio galaxies and radio-loud quasars.  The
radio-quiet objects follow a remarkably tight correlation between
their 60 \micron\ far-infrared (FIR) and GHz radio continuum
luminosities \citep{kru73, ric84, dic84, jon85, hel85, con86, wun87,
con88, wal89, con91a, sop91, cra92}, although Seyfert galaxies and
radio-quiet quasars appear to scatter somewhat more than starburst
galaxies \citep{san85, con88, bau93}.  The correlation is attributed
to stellar processes that are thought ultimately to be responsible for
both the radio and the FIR emission.  Radio-loud objects lie up to
$\sim$ 3 dex in radio luminosity above the starburst line. Their radio
emission is clearly associated with an active nucleus. A radio/FIR
correlation has been suggested for radio-loud objects, but with
significantly more scatter than for radio-quiet objects \citep{sop91}.
The lower edge of this relation, at least at 5 GHz, corresponds
approximately with the equivalent flux limit of the 408 MHz revised 3C
radio catalog \citep{ben63}, suggesting that it may be an artifact of
the sample selection rather than having a physical basis.

It has recently become apparent that a class of objects exists with
intermediate radio properties. \citet{dey94} correlated the 60
\micron\ \iras\ Faint Source Catalog (FSC) with the Texas 365 MHz
radio survey and identified objects lying at intermediate radio
luminosities in the radio/FIR diagram.  Many of these objects have
optical spectra with features characterized as post-starburst AGN;
strong Balmer absorption lines and Seyfert 2 emission lines ([O~III]
$\lambda$5007 stronger than H$\alpha$) or LINER-like emission lines
([O~II] $\lambda$3727 stronger than [O~III] $\lambda$5007, [O~I]
$\lambda$6300/[O~III] $\lambda$5007 $>$ 0.3; Heckman 1980).  Condon,
Anderson, \& Broderick (1995, hereafter CAB) correlated the 60
\micron\ \iras\ FSC with the Greenbank 4.8 GHz radio survey and
identified a similar sample of radio-excess galaxies, although they
did not explicitly comment on this.  The radio-intermediate objects in
their sample encompass a range of FIR luminosities\footnote{The FIR
luminosity is estimated from the FIR flux density
$S_{FIR}=1.26{\times}10^{-14} (2.58 \Snfir + \Snhun) \> $
Wm$^{-2}$ \citep{san96}.} from low luminosity systems with $L_{FIR}
\sim 10^{8} \> L_\odot$ to highly luminous infrared galaxies with
$L_{FIR} \geq 10^{11} \> L_\odot$.  Subsequently, \citet{roy97} showed
that selecting FIR luminous galaxies that have a radio excess is an
effective method of finding dusty, gas-rich active galactic nuclei
(AGN). More recently, Yun, Reddy, \& Condon (2001) correlated \iras\
FSC sources above 2 Jy at 60 \micron\ with the NRAO \vla\ Sky Survey
\citep[NVSS]{con98} at 1.4 GHz, and identified a small number of
radio-intermediate objects (23 of 1809).  \citet{cor02} have also
identified a small number of radio-intermediate objects in their
sample of low power AGN.

Ultra-luminous infrared galaxies (ULIRGs) feature among the most FIR
luminous radio-intermediate objects\footnote{Luminous infrared
galaxies are defined formally to have $10^{11} < L_{FIR} < 10^{12} \>
\Lsun$ and ultra-luminous infrared galaxies are defined to have
$L_{FIR} > 10^{12} \> \Lsun$ \citep{san96}.}. Most ultra-luminous
infrared galaxies are in interacting or recently merged systems
\citep{soi87}, and include both starburst galaxies and AGN.  Indeed,
both types of activity coexist in many ultra-luminous infrared
galaxies. This supports the idea that galaxy interactions trigger
intense nuclear starbursts that may evolve into AGN activity
\citep{san88a, san88b}.  Multiple nuclei are common, suggesting that
the progenitors of many ultra-luminous infrared galaxies may be
compact groups of galaxies \citep{bor00}.  Radio-intermediate
infrared-luminous galaxies share some of these characteristics and so
may be relevant to studies of how AGN activity is initiated.

The radio emission from radio-intermediate ultra-luminous infrared
galaxies may arise in several ways.  At one extreme, they may be
``weak'' radio galaxies that will never achieve radio luminosities
commensurate with those of 3C radio galaxies and quasars.  At the
other extreme, they may be dusty, gas-rich systems in which the radio
activity is not yet fully developed because the nascent radio jet has
only just started to emerge in the confining interstellar medium (ISM)
of the host galaxy.  Radio-intermediate ultra-luminous infrared
galaxies may therefore share some similarities with compact
steep-spectrum (CSS) and GigaHertz peaked-spectrum (GPS) radio sources
\citep{ode98}. CSS radio sources are objects in which the radio lobes
are contained within the $\sim$ 15 kpc extent of the host galaxy and
the radio emission is optically thin, unlike the flat-spectrum cores
of radio galaxies and radio-loud quasars \citep{fan90, you93, gel94,
fan95}. GPS sources are believed to be similar to CSS objects, but
with scales of $<$ 1 kpc.  CSS and GPS sources have steep radio
spectra at high radio frequencies, but the spectra of GPS sources turn
over around 1 GHz and decline with decreasing frequency, whereas the
spectra of CSS sources continue to rise \citep{ode90a, ode90b, ode91,
ode96}.  This turnover may be due to free-free absorption (FFA) in
material surrounding the radio lobes (e.g., Bicknell, Dopita, \& O'Dea
1997) or to synchrotron self-absorption (SSA) in the emitting plasma
(e.g., Readhead et al. 1996).  CSS/GPS radio sources may be young
\citep{fan95}, or the interstellar medium of the host galaxy may be
sufficiently dense to confine the radio plasma \citep{bre83}.

In this paper, we define a sample of radio-excess infrared galaxies
based on source detections in the 60 \micron\ \iras\ FSC and the
Parkes-MIT-NRAO (PMN) 4.8 GHz point source catalog.  This sample is
combined with similar CAB objects in an attempt to better determine
the nature of radio-excess objects.  Our cross-correlation of the
catalogs and subsequent observations are described in \S\ref{sec:ids}.
Results for the full sample of radio-quiet and radio-excess objects
are presented in \S\ref{sec:res} where we also define a radio-excess
sub-sample. The nature of these radio-excess objects is explored in
\S\ref{sec:kno} by noting the many well-studied objects in the
sample. The properties of the radio-excess sample as a whole are
considered in \S\ref{sec:dis} where we highlight similarities with
CSS/GPS sources.  Our conclusions are summarized in \S\ref{sec:con}.
We adopt $H_{0}$ = 50 km~s$^{-1}$~Mpc$^{-1}$ and $q_0$ = 0.5
throughout.

\section{Radio-\iras\ Identifications}
\label{sec:ids}

\subsection{Cross-Correlations}
\label{sec:ids_cor}

The PMN/FSC sample is based on a cross-correlation of the
Parkes-MIT-NRAO (PMN) 4.8 GHz catalog with 60 \micron\ detections in
the \iras\ FSC.  The PMN survey covered the whole sky between
declinations +10$^\circ$ and -87$^\circ$ to a 4.8 GHz flux density
limit of typically $\sim$ 30 mJy, rising to $\sim$ 45 mJy in
equatorial regions \citep{gri93, gri94, wri94, gri95, wri96}. The
catalog derived from these data has a positional accuracy of $\sim$
10\arcsec\ for bright sources, and falls to $\sim$ 40\arcsec\ at the
flux density limit of the survey \citep{gri93}.  The \iras\ FSC
\citep{mos92} at 60 \micron\ covers 94\% of the sky with $|b| \geq
10^\circ$ to a flux density limit of $\sim$ 200 mJy and a positional
uncertainty typically of $\sim$ 7\arcsec\ $\times$ 25\arcsec.  We
require an \iras\ detection only at 60 \micron\ for inclusion in our
sample. Hence we have considered more candidate objects than previous
samples that have been based on \iras\ color selection (e.g., CAB; Roy
\& Norris 1997; Yun et al. 2001). This leads to the inclusion of a
larger number of faint \iras\ detections in our sample.

The cross-correlations were performed using the XCATSCAN facility at
the Infrared Processing and Analysis Center\footnote{IPAC is operated
by the Jet Propulsion Laboratory, California Institute of Technology,
under contract with the National Aeronautics and Space
Administration.} (IPAC). Initially, all sources with PMN positions
within 60\arcsec\ of an \iras\ position were accepted.  Correlations
with Galactic objects such as stars, star clusters, planetary nebulae,
and \ion{H}{2} regions were then removed, as were correlations in the
direction of the Magellanic Clouds. The resulting sample consisted of
279 radio/FIR correlations.  The NASA/IPAC Extragalactic
Database\footnote{NED is operated by the Jet Propulsion laboratory,
California Institute of Technology, under contract with the National
Aeronautics and Space Administration} (NED) was then searched for data
on each correlation.  Many correlations correspond to bright galaxies
for which the radio/FIR association is unquestioned.  Others
correspond to fainter galaxies or quasars with previously measured
redshifts, while the remainder either correspond to objects with
unknown redshifts that are faint on Digitized Sky Survey\footnote{DSS,
The Digitized Sky Survey was produced at the Space Telescope Science
Institute under U.S. Government grant NAG W-2166. The images of these
surveys are based on photographic data obtained using the Oschin
Schmidt Telescope on Palomar Mountain and the UK Schmidt Telescope.}
(DSS) images or for which the optical identification is confused.

\subsection{Valid Correlations}
\label{sec:ids_ass}

We now consider whether each of the 279 correlations is a valid
association of the FIR and radio sources.  In general, the PMN radio
positions are not of sufficient accuracy to allow us to make
unambiguous associations with FIR sources or to assign optical
identifications based solely on positional coincidence.  We therefore
used information from the literature and new observations to improve
the radio positional accuracy for each correlation. We adopt as our
criterion for acceptance of a radio/FIR correlation in the final
sample that the accurate (i.e., $<$ 5\arcsec) radio position lies
within the \iras\ FSC 60 \micron\ 3$\sigma$ error ellipse.

A number of more accurate radio positions were available from the
literature.  CAB have performed a similar selection in the northern
hemisphere in the region 0$^h < \alpha < 20^h$ and $\delta >
-40^\circ$.  They list 4.8 GHz positions with $\sim$ 1\arcsec\
uncertainty measured from \vla\ snapshots for 58 of our radio/FIR
correlations.  Accurate 1.4 GHz radio positions ($\sim$ 1\arcsec) are
also available from the NVSS \citep{con98} for a further 127 of our
correlations north of declination $-40^\circ$.  The NVSS data confirm
116 of these radio/FIR correlations, while 11 correlations were
rejected on the basis that the nearest 1.4 GHz radio source lies
outside the \iras\ $3\sigma$ error ellipse.

Snapshots at 3 cm and 6 cm were obtained with the Australia Telescope
Compact Array (\atca) for 54 southern correlations. Observations were
made in 1996 July and 1998 January using the 6C configuration and in
1998 February using the 6B configuration. Both configurations have a
maximum baseline of 6 km. Antenna 2 was out of operation during the
1998 January observing run, resulting in the loss of 5 baselines,
including the shortest one (153 m). Sources were observed at 8.64 GHz
and 4.79 GHz with a bandwidth of 128 MHz.  Unlike the \vla, the \atca\
is a linear E-W array of six 22 m dishes so several observations of
each object at different hour angles (i.e., cuts) were required to
minimally sample the $u-v$ plane.  Generally 6 cuts of 3 minutes each
were obtained for each object, evenly spaced in time while the object
was accessible above the horizon.  A phase calibrator was observed
within $\sim$~20 minutes in time and $\sim$ 10$^{\circ}$ in angle on
the sky of each group of 1--6 target sources. The \atca\ primary
calibrator, PKS B1934-638, was observed during each session to
calibrate flux densities.

The data were reduced using \miriad\ \citep{sau95}. Bad data were
flagged and data from all observing dates were combined. MFCAL was
used to flux calibrate the visibilities relative to PKS B1934-638,
which is assumed to have flux densities of 2.842 Jy and 5.843 Jy at
8.64 GHz and 4.79 GHz, respectively. Phase solutions were calculated
for the secondary calibrators using GPCAL then phase calibration was
applied to the target sources using GPCOPY. Maps of target sources
were made and cleaned using MFCLEAN. Cleaned maps were restored using
a Gaussian beam with uniform weighting. All but two sources were
self-calibrated; self-calibration was omitted for one source because
it is too faint and for another because it is too extended.  Total
integrated fluxes were measured using IMFIT in \miriad.  These are
listed in Table \ref{tab:sam}.  Flux density limits ($3\sigma$) at the
two frequencies are typically $\sim$ 1.7 mJy~beam$^{-1}$ and $\sim$
1.2 mJy~beam$^{-1}$, respectively. The beam is approximately circular
for most sources, with a FWHM of $\sim$ 1\arcsec\ at 8.64 GHz and
$\sim$ 2\arcsec\ at 4.79 GHz. For sources approaching the northern
limit of the \atca\ of $\sim -10^{\circ}$ declination, the beam is
progressively more elongated in the North-South direction up to $\sim$
90\arcsec\ in extreme cases. The errors in the measured source
positions are estimated to be $\sim$ 1--5\arcsec\ with the larger
values in this range applying to the more northern sources that have
elongated beams. This positional accuracy is adequate for confirming
radio/FIR correlations and assigning optical identifications in
unconfused regions.
 
The \atca\ observations confirmed 28 of the radio/FIR correlations,
while the radio source was either not detected or was deemed not to be
associated with the FIR source in a further 26 correlations.

Optical counterparts to the radio sources were then sought on
Digitized Sky Survey images. In many cases, only one optical object is
present in the \iras\ 3$\sigma$ error ellipse and this is confirmed to
be the radio object. However, when more than one optical object is
present in the \iras\ 3$\sigma$ ellipse, this procedure leaves
uncertain whether the radio/optical object is also the FIR source.
These cases were deemed to be uncertain FIR identifications and were
rejected. A total of 24 correlations were excluded due to this optical
confusion.

The final sample of 178 objects are confident associations of PMN
radio sources with \iras\ FSC 60 \micron\ sources.  These 178 objects
are listed in Table \ref{tab:sam}. Column 1 lists the \iras\ name.
Column 2 lists the PMN name. Columns 3 and 4 list the J2000 source
position. Column 5 indicates the origin of the source position;
positions from \citet{con95} are listed as CAB, positions from the
NVSS are listed as NVSS, positions from our \atca\ snapshots are listed
as ATCA, PMN positions are listed as PMN and positions derived from
the optical host galaxies measured from DSS images are listed as DSS.
Columns 6 and 7 list the \iras\ 25 \micron\ and 60 \micron\ flux
densities; except where noted, fluxes are from the FSC. Columns 8 and
9 list the 8 GHz and 5 GHz flux densities, respectively. Column 10
lists the origin of the 5 GHz flux density as either our \atca\ 
snapshots or the PMN catalog. The single-dish PMN measurements suffer
from source confusion so their flux densities at low levels can be
unreliable. Columns 11 and 12 list redshifts for the optical
counterparts and their origins (see \S\ref{sec:ids_red}).  Column 13
lists alternative designations for the objects derived from NED.

\placetable{tab:sam}
\begin{deluxetable}{llrrcrrrrcrcl}
\tabletypesize{\scriptsize}
\tablewidth{\textwidth}
\tablewidth{0pt}
\tablecolumns{13}
\tablecaption{PMN/FSC Sample\label{tab:sam}}
\rotate
\tablehead{
\colhead{FSC Name} &
\colhead{PMN Name} &
\multicolumn{2}{c}{Position\tablenotemark{a}} &
\colhead{Ref.} &
\colhead{$S_{\nu}$(25 \micron)} &
\colhead{$S_{\nu}$(60 \micron)} &
\colhead{$S_{\nu}$(8 GHz)} &
\colhead{$S_{\nu}$(5 GHz)} &
\colhead{Ref.} &
\colhead{$z$} &
\colhead{Ref.} &
\colhead{Identification} \\
& &
\multicolumn{2}{c}{RA\phn(J2000)\phn DEC} 
& &
\colhead{(Jy)} &
\colhead{(Jy)} &
\colhead{(mJy)} &
\colhead{(mJy)} 
\\
\colhead{(1)} &
\colhead{(2)} &
\colhead{(3)} &
\colhead{(4)} &
\colhead{(5)} &
\colhead{(6)} &
\colhead{(7)} &
\colhead{(8)} &
\colhead{(9)} &
\colhead{(10)} &
\colhead{(11)} &
\colhead{(12)} &
\colhead{(13)}
}
\startdata
F00123-2356 & PMNJ0014-2339  & 00 14 52.7  & -23 39 20 & NVSS &$<$0.085\phn\phn\phn& 0.225\phn\phn\phn& 20.0\phn\phn\phn& 53.8\phn\phn\phn&    ATCA & 0.0830 &    R150 & Abell~14        \\
F00124-3929 \tablenotemark{b}
            & PMNJ0014-3912  & 00 14 58.5  & -39 12 29 &  CAB & 6.250\phn\phn\phn& 77.000\phn\phn\phn& \nodata\phn\phn& 112.0\phn\phn\phn&     CAB & 0.0005 &     NED & NGC 55          \\
F00183-7111 \tablenotemark{c}
            & PMNJ0020-7055  & 00 20 34.7  & -70 55 26 & ATCA & 0.133\phn\phn\phn& 1.195\phn\phn\phn& 51.9\phn\phn\phn& 108.6\phn\phn\phn&    ATCA & 0.3270 &     DBS & IRAS~00182-7112 \\
F00402-2349 & PMNJ0042-2333  & 00 42 45.8  & -23 33 41 & NVSS & 1.077\phn\phn\phn& 10.040\phn\phn\phn& \nodata\phn\phn& 56.0\phn\phn\phn&     PMN & 0.0202 &     NED & NGC 232         \\
F00450-2533 \tablenotemark{b}
            & PMNJ0047-2517  & 00 47 33.0  & -25 17 18 &  CAB & 155.650\phn\phn\phn& 998.730\phn\phn\phn& \nodata\phn\phn& 2980.0\phn\phn\phn&     CAB & 0.0007 &     NED & NGC 253         \\
F00573-0750 & PMNJ0059-0734  & 00 59 50.3  & -07 34 44 & NVSS & 0.647\phn\phn\phn& 8.347\phn\phn\phn& \nodata\phn\phn& 44.0\phn\phn\phn&     PMN & 0.0055 &     NED & NGC 337         \\
F01009-3241 & PMNJ0103-3225  & 01 03 16.7  & -32 25 56 &  CAB &$<$0.123\phn\phn\phn& 0.291\phn\phn\phn& 36.8\phn\phn\phn& 95.3\phn\phn\phn&    ATCA & 0.2560 &     DBS & PKS~0100-326    \\
F01044-4050 & PMNJ0106-4034  & 01 06 45.1  & -40 34 20 & ATCA & 0.155\phn\phn\phn& 0.405\phn\phn\phn& 4414.0\phn\phn\phn& 3377.0\phn\phn\phn&    ATCA & 0.5840 &     NED & Q0104-408       \\
\enddata
\protect\tiny{
\tablenotetext{a}{Right Ascension is given in units hours, minutes and seconds. Declination is given in units of degrees, arcminutes and arcseconds.} 
\tablenotetext{b}{FIR fluxes are from \citet{ric88}.} 
\tablenotetext{c}{Radio/FIR association discussed by \citet{nor88}.} 
\tablenotetext{d}{FIR fluxes are from \citet{soi89}.} 
\tablenotetext{e}{Varied by factor of 1.85 at 8.6 GHz between July 1996 (107.7 mJy) and Jan/Feb 1998 (199.3 mJy).} 
\tablenotetext{f}{ATCA flux is fainter than CAB ($S_\nu(4.8 GHz) = 188$ mJy) and PMN ($S_\nu(4.8 GHz) = 172.0$ mJy).} 
}
\end{deluxetable}

\subsection{Reliability of Identifications}
\label{sec:ids_rel}

A sample selected by positional coincidence alone may contain spurious
associations due to the chance presence of a PMN source within the
\iras\ 3$\sigma$ error ellipse. We have estimated the number of chance
coincidences in our sample from both the average PMN source density
and by offsetting the positions and recalculating the
cross-correlation.

The number of chance coincidences can be estimated from the product of
the average PMN source density in the overlap area between the PMN and
FSC surveys and the total sky area contained within all \iras\ 
3$\sigma$ error ellipses. The survey overlap area is 6.13 steradians,
which is bounded by declination +10$^\circ$ and -87.5$^\circ$ and
excludes the Galactic plane in the region $|b| < 10^\circ$ and a
further area of $1.8 \times 10^{-2}$ steradians due to the Large and
Small Magellanic Clouds. The PMN catalog contains 38,170 sources within
this survey overlap area, corresponding to an average PMN source
density of $6.2 \times 10^3$ sr$^{-1}$. A total sky area of $3.85
\times 10^{-3}$ steradians is contained within the 3$\sigma$ error
ellipses of all candidate 60 \micron\ \iras\ FSC sources searched.
This includes the 3$\sigma$ error ellipses of both the FSC 60 \micron\ 
sources with radio counterparts and those without counterparts. We
therefore expect 23.9 chance coincidences of PMN sources within
unrelated FSC error ellipses.

A more accurate method is to offset the PMN source positions and to
recalculate the correlations. Offsetting the PMN positions further
than the extent of the largest \iras\ error ellipse ensures that all
derived correlations are spurious. Offsetting by an amount not much
larger than this ensures that the correlations are recalculated with
close to the true PMN source density in the vicinity of each FSC
source. We chose to offset the PMN positions 2 arcmin north. A total
of 25 spurious correlations were then obtained. We are therefore
confident that the number of chance coincidences included in our full
sample is $\sim$ 24-25.

The full sample contains 243 valid PMN/FSC correlations comprising 178
extragalactic objects with optical counterparts, 24 extragalactic
correlations that were rejected on the basis of optical confusion, and
41 Galactic objects. In all, 17\% of the valid correlations are
identified with Galactic objects. The same percentage of Galactic
objects should be present among the $\sim$ 24-25 expected chance
coincidences, so our extragalactic correlations alone should include
$\sim$ 21 chance coincidences. This is consistent with the 24
extragalactic correlations that in fact were rejected on the basis of
optical confusion.

Consequently, we are confident that the final list of 178 PMN/FSC
identifications are reliable and that chance coincidences in our
sample have been identified and excluded by reference to the optical
images.

\subsection{Redshifts}
\label{sec:ids_red}

Redshifts for the optical counterparts of 143 of the PMN/FSC objects
are available in the NED database. Redshifts for some other objects have
been derived from the proper distances tabulated from various sources
by CAB.

Redshifts for most of the remaining objects were derived from optical
spectra obtained with the Double Beam Spectrograph \citep[DBS]{rod88}
on the Australian National University (ANU) 2.3 m telescope at Siding
Spring Observatory. The B300 and R316 gratings were used in the blue
and red arms, respectively, with a 2\arcsec\ slit giving full
wavelength coverage from $\sim$ 3500 \AA\ to $\sim$ 9000 \AA\ with a
resolution of $\sim$ 4 \AA.  Two dichroic mirrors were used for
different redshift regimes to avoid locating prominent emission lines
near the cutoff wavelength; dichroic \#3 has a cut-off wavelength of
6200 \AA\ so was used for $z < 0.20$ and $z > 0.32$, and dichroic \#1
has a cutoff wavelength of 5630 \AA\ so was used for $0.20 < z <
0.32$.  Each observation consisted of three unguided exposures each of
1200 s duration.  Object acquisition was by offsetting from a nearby
star before each 1200 s exposure. Telescope tracking was adequate to
hold the object within the 2\arcsec\ slit over this time.  The data
were reduced using the \onedspec\ package within \iraf\footnote{\iraf\
(\emph{Image Reduction and Analysis Facility}) is distributed by the
National Optical Astronomy Observatories, which are operated by the
Association of Universities for Research in Astronomy, Inc., under
cooperative agreement with the National Science Foundation.}.
Generally, a 9\arcsec\ region around each object was extracted from
the two dimensional image.  Wavelength calibration was based on a
single Fe-Ar arc spectrum obtained each night.  Wavelength shifts
during each night were tracked using night sky emission lines recorded
on each exposure.

Some redshifts were determined using the Nasmyth B Imager
\citep{rod93} on the ANU 2.3~m telescope.  Spectra were measured
through either the R150 or R300 grisms with a 2.5\arcsec\ wide
slit. The R150 grism delivered a resolution of $\sim$ 44 \AA\ over a
wavelength range of $\sim$ 4000--10000 \AA. The R300 grism delivered a
resolution of $\sim$ 22 \AA\ over a wavelength range of typically
6000--9000 \AA. Unguided exposures of 1800~s were obtained, and He and
Ar arc spectra were recorded for wavelength calibration. The target
acquisition and data reduction procedures were the same as described
above for the DBS spectra.

For all object spectra obtained at the 2.3 m telescope, measurements
of smooth-spectrum stars obtained at similar airmasses were used to
remove terrestrial absorption features. Flux calibration was performed
relative to standard stars from either \citet{oke83} or \citet{ham92,
ham94}.  Emission-line redshifts were measured using Gaussian line
profile fits in \splot. The redshifts are listed in Table
\ref{tab:sam}.

\section{Results}
\label{sec:res}

\subsection{Radio/Far-infrared Correlations}
\label{sec:res_rad}

\begin{figure}
\plottwo{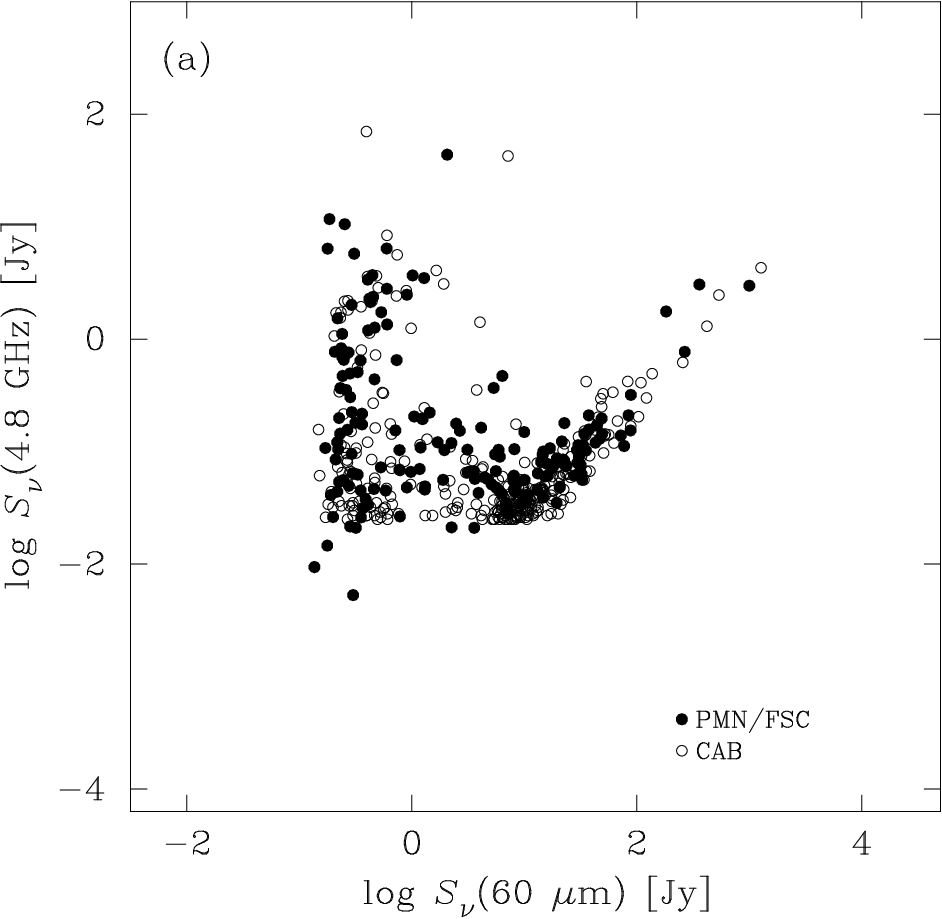}{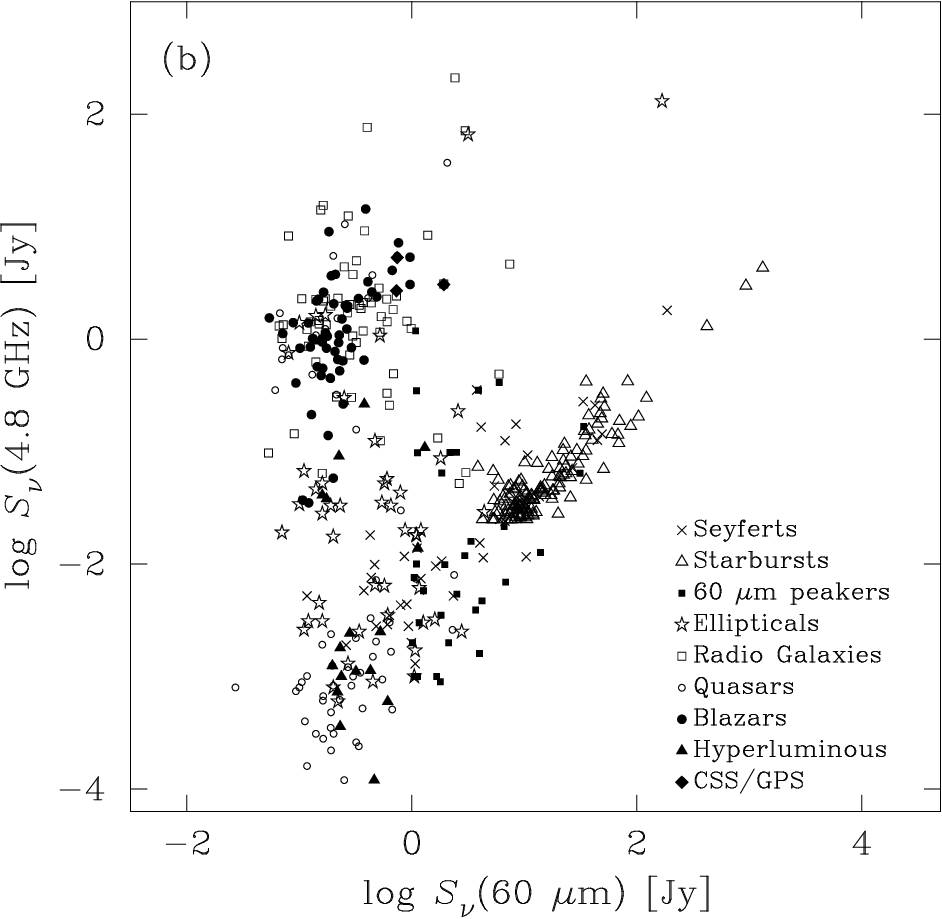} 
\caption{(a) Radio/FIR flux density diagram for the PMN/FSC sample
  (\emph{solid symbols}) and the CAB sample (\emph{open symbols}).
  (b) Radio/FIR flux density diagram for the starburst and active
  galaxy samples listed in Table \ref{tab:com}.\label{fig:fluxes}}
\end{figure}

The radio/FIR flux density diagram for the PMN/FSC sample is shown in
Fig.\ \ref{fig:fluxes}(a).  The tight sequence of FIR-bright objects
is a manifestation of the well-known radio/FIR correlation for
radio-quiet objects.  Of particular note is the large number of
relatively FIR-faint radio-excess objects located above the
radio-quiet relation. These objects are also present in the sample of
CAB (open symbols in Fig.\ \ref{fig:fluxes}(a)) but were not
specifically noted by these authors. Their sample has a similar radio
flux density to the PMN/FSC sample and both were selected from the
\iras\ FSC (60 \micron\ flux limit of $\sim$ 200 mJy), but CAB imposed
the additional criterion that $S_\nu$(60 \micron) $>$ $S_\nu$(12
\micron).  Only a few radio-excess objects are present in other
samples with higher FIR flux density limits (e.g., \citet{yun01} with
a 60 \micron\ flux limit of 2 Jy and \citet{cor02} with a 60 \micron\
flux limit of 4 Jy). A large number of the radio-excess objects appear
at 60 \micron\ fluxes below 1 Jy, with many of these having
intermediate radio excesses. We refer to these latter objects as
radio-intermediate objects. The analysis presented in
\S\ref{sec:ids_rel} demonstrates the reliability of our sample
identifications. The FIR-faint radio-intermediate objects therefore
appear to be a new population of extragalactic objects.

\placefigure{fig:fluxes}

\placetable{tab:com}
\begin{deluxetable}{lll}
\tabletypesize{\scriptsize}
\tablewidth{\textwidth}
\tablecaption{Comparison Objects\label{tab:com}}
\tablehead{
\colhead{Class} &
\colhead{Far-Infrared Data} &
\colhead{Radio Data}
}
\startdata
Seyfert galaxies            & \citet{ede87b}   & \citet{ede87a} \\
Infrared galaxies           & \citet{soi89}    & Condon, Frayer, \& Broderick (1991b) \\
\nodata                     & \nodata          & CAB \\
60~$\mu$m Peakers           & \citet{vad93}    & \citet{hei97}  \\
Nearby ellipticals          & \citet{kna89}    & \citet{sad84}  \\
\nodata                     & \nodata          & \citet{sad89}  \\
Radio galaxies              & \citet{gol88}    & \citet{kuh81}  \\
\nodata                     & \citet{kna90}    & \nodata        \\
Quasars                     & \citet{san89}    & \citet{kel89}  \\
\nodata                     & \citet{elv94}    & \citet{elv94}  \\
Blazars                     & \citet{imp88}    & \citet{owe80}  \\
\nodata                     & \citet{lan86}    & \citet{lan86}  \\
Hyperluminous galaxies      & \citet{row00}    & \citet{bar89,bar97}\\
\nodata                     & \nodata          & \citet{blu95,gre91} \\
\nodata                     & \nodata          & \citet{gri93,kel89} \\
\nodata                     & \nodata          & \citet{row00} \\
\nodata                     & \nodata          & \citet{bec95,con98} \\
\nodata                     & \nodata          & (based on $S_\nu$(1.4 GHz) with $\alpha = -0.8$) \\
CSS and GPS sources         & \iras\ FSC       & \citet{ode98} 
\enddata
\end{deluxetable}

It is useful to consider the locations of well-studied objects in the
radio/FIR flux density diagram in order to determine what types of
objects populate the radio-intermediate region of Fig.\
\ref{fig:fluxes}.  The positions of a range of well-studied active and
starburst galaxies in the radio/FIR flux density diagram are shown in
Fig.\ \ref{fig:fluxes}(b). Table \ref{tab:com} lists the references to
these data.  The relative densities of these objects in Fig.\
\ref{fig:fluxes}(b) are not indicative of unbiased samples because the
objects are drawn from heterogeneous samples with different flux
density limits. Nevertheless, Fig.\ \ref{fig:fluxes}(b) is still
useful in highlighting where different classes of objects lie.  As is
well-known, starburst galaxies, Seyfert galaxies, and radio-quiet
quasars generally follow the radio-quiet correlation in Fig.\
\ref{fig:fluxes}, with some Seyfert galaxies lying significantly above
the radio-quiet sequence.  3C radio galaxies and radio-loud quasars
follow the apparent radio-loud equivalent of this relation. This is a
selection effect due to the high 3C flux density limit: The 3C flux
limit of $\sim$ 10 Jy at 408 MHz \citep{ben63} for a steep spectrum
object with radio spectral index of 0.7 corresponds to $\sim$ 1.7 Jy
at 5 GHz. This is close to the lower edge of the apparent radio-loud
clump of objects in Fig.\ \ref{fig:fluxes}(b).  Interestingly, many of
the nearby elliptical galaxies surveyed for radio emission by
\citet{sad84} have radio and FIR flux densities that place them in
the radio-intermediate region of Fig.\ \ref{fig:fluxes}(b). Some
well-studied radio galaxies, Seyfert galaxies, and 60 $\mu$m Peakers
also extend into this region. These are discussed in more detail in
\S\ref{sec:kno}.

\begin{figure}[ht]
\plottwo{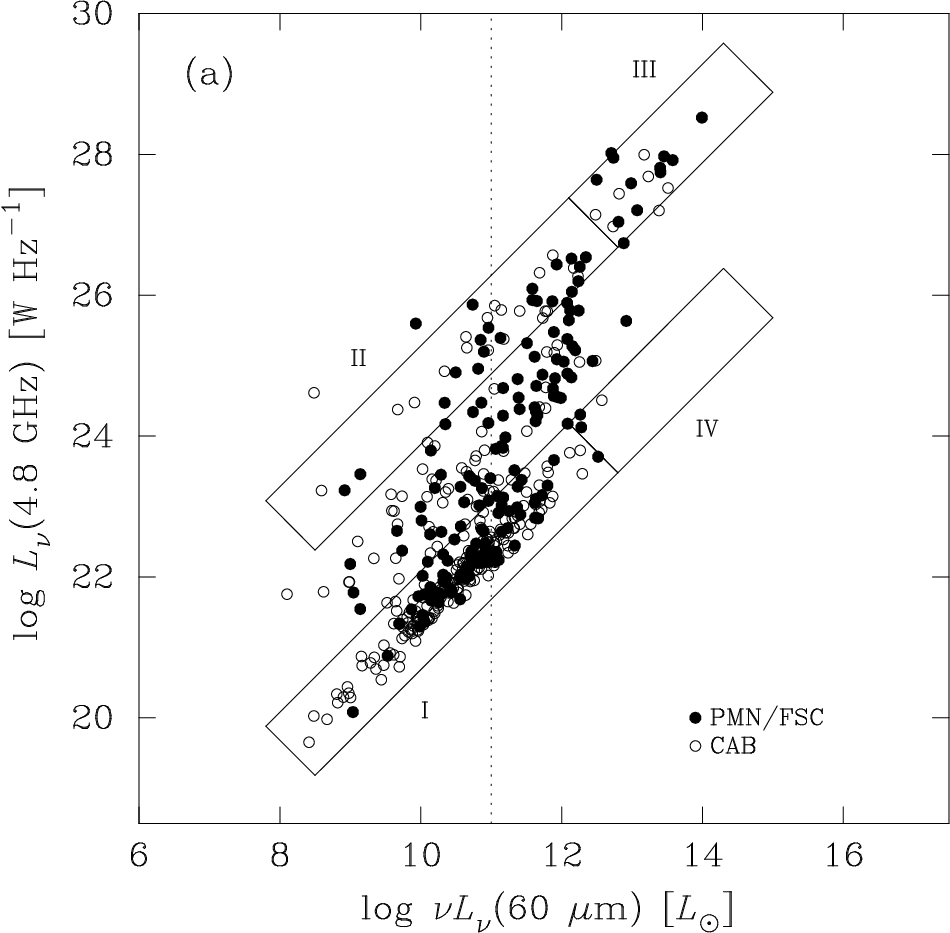}{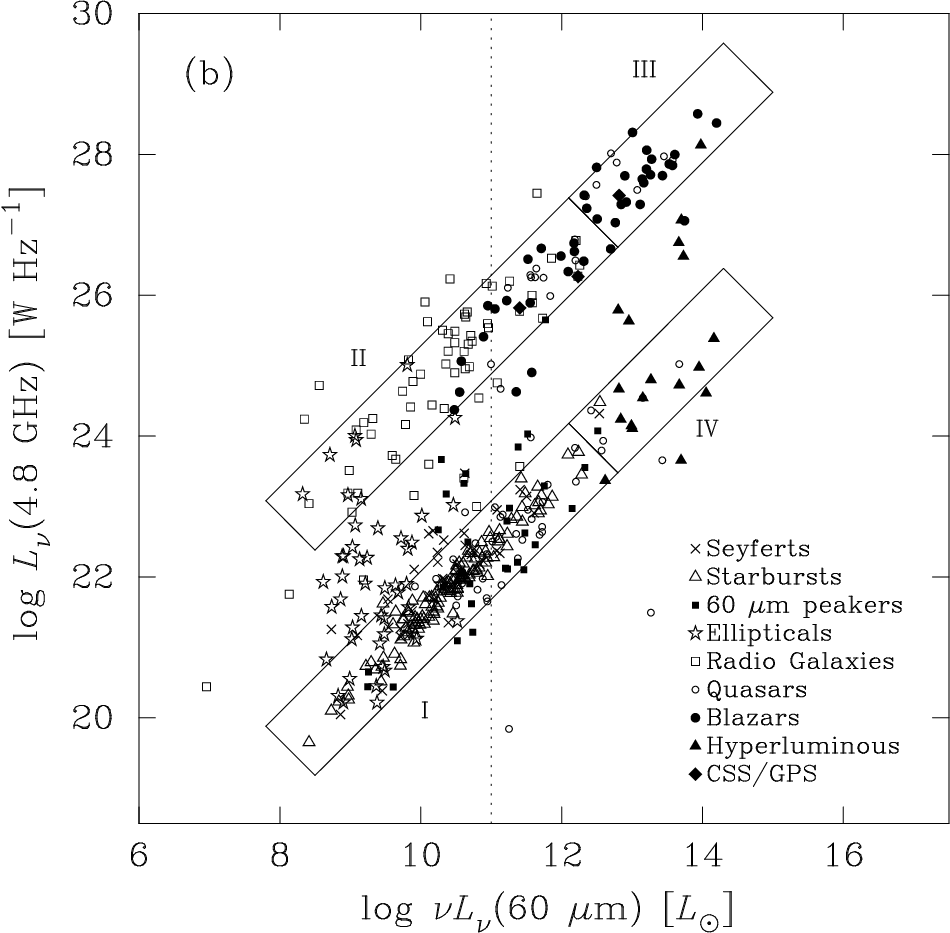} 
\caption{(a) Radio/FIR luminosity diagram for the PMN/FSC sample
  (\emph{solid symbols}) and the CAB sample (\emph{open symbols}).
  The vertical dotted line at $\nLnfir = 10^{11} \> \Lsun$ separates
  the low and high FIR luminosity samples.  The four rectangular
  regions are discussed in the text. 
  (b) Radio/FIR luminosity diagram for the starburst and active galaxy
  samples listed in Table \ref{tab:com}.\label{fig:lums}}
\end{figure}

The corresponding radio/FIR luminosity diagram for the PMN/FSC sample
is shown in Fig.\ \ref{fig:lums}(a).  Distances have been derived from
the redshifts listed in Table \ref{tab:sam}. The CAB objects are also
plotted in Fig.\ \ref{fig:lums}(a) with distances taken from that
paper.  Specific luminosity (i.e., $L_\nu$) is plotted for the radio
data and monochromatic luminosity (i.e., $\nu L_\nu$) is plotted for
the FIR data.  There are two points to note from this figure: First,
the radio-excess objects extend over the full range of 60 \micron\
luminosities shown by the radio-quiet objects (from $\sim 10^8 \>
\Lsun$ to $\gtrsim 10^{12.5} \> \Lsun$). The high end is in the range
of ultra-luminous infrared galaxies. In fact, the radio-loud objects
extend to even higher FIR luminosities than radio-quiet objects. There
is possibly a lack of low FIR luminosity radio-excess objects in the
PMN/FSC sample.  However, this may be due to nearby lobe-dominated
radio galaxies being excluded from the PMN/FSC sample due to large
position offsets between their radio (lobe) and FIR (host) emission.
Second, there is no obvious radio-loud equivalent of the radio-quiet
radio/FIR luminosity correlation in these samples.  In fact, the
radio-excess region appears to be populated uniformly. The
distribution of radio-excess objects is accurately represented in
Fig.\ \ref{fig:lums}(a) at high radio excesses. However, objects with
small radio excesses are under-represented in Fig.\ \ref{fig:lums}(a)
because objects are missing that fall below the $\sim$ 30 mJy flux
density limits of both the PMN and Greenbank radio surveys. This means
that the true distribution of radio-excess objects is more densely
populated at low radio excesses close to the radio/FIR correlation
than is indicated by Fig.\ \ref{fig:lums}(a).

\placefigure{fig:lums}

\clearpage

The comparison objects shown in Fig.\ \ref{fig:fluxes}(b) are
replotted in the radio/FIR luminosity diagram in Fig.\
\ref{fig:lums}(b).  We identify four regions in Fig.\
\ref{fig:lums}(b) based on the locations of these objects; i) the
sequence of radio-quiet objects on the radio/FIR correlation, ii) the
apparent sequence of 3C radio galaxies and radio-loud quasars, iii) an
extension of the radio-loud sequence to \nLnfir\ $\gtrsim 10^{12} \>
\Lsun$ that is occupied predominantly by blazars, and iv)
hyperluminous \iras\ galaxies forming an extension of the radio-quiet
sequence to \nLnfir\ $\gtrsim 10^{12} \> \Lsun$.  The dominant source
of both the radio and FIR luminosity from starburst objects in region
i) is believed to be related to star formation. This group also
includes radio-quiet quasi-stellar objects (QSO) for which there is
less consensus on the origin of the radio and FIR luminosity. As
mentioned above, the apparent correlation of objects in region ii)
parallel to the radio-quiet correlation is likely to be due to the
high flux density limit of the 3C catalog, rather than to a real
correlation between the radio and FIR luminosities. AGN dominate the
radio luminosity from these objects.  The radio emission from the
blazars in region iii) is believed to be relativistically beamed
towards Earth, so isotropic luminosities over-estimate the true
luminosity. The FIR emission from these objects may also be
non-thermal and possibly beamed.  The hyperluminous galaxies in region
iv) include \iras\ F10214+4724 and the Cloverleaf quasar, both of
which are gravitationally lensed \citep{bro95, mag88}. These objects
are rare, either because of their extreme intrinsic luminosities or
because gravitational lensing of the required magnification is rare.
Indeed, no objects are present in this region of Fig.\
\ref{fig:lums}(a). Radio-intermediate objects are represented in Fig.\
\ref{fig:lums}(b), but only in significant numbers at luminosities
below $\nLnfir \sim 10^{11.5} \> \Lsun$. Again, it is clear that only
the \citet{sad84} nearby elliptical galaxies significantly populate
the radio-intermediate region, with a few radio galaxies and Seyfert
galaxies also extending into this region.  The nearby elliptical
galaxies are generally low luminosity objects with $\nLnfir <
10^{10.5} \> \Lsun$. There are few radio-intermediate objects in the
comparison samples with $\nLnfir \gtrsim 10^{11.5} \> \Lsun$.  This
reinforces the suggestion that the high FIR luminosity
radio-intermediate objects in Fig.\ \ref{fig:lums}(a) are a previously
unrecognized population.  Since most of the radio-excess objects have
similar 60 \micron\ flux densities slightly above the FSC limit, the
high luminosity examples must all be at significant distances; an
object with $\Snfir < 1$ Jy and $\nLnfir > 10^{11.5} \> \Lsun$ must be
at redshift $z > 0.075$ in the adopted cosmology. These moderate
redshift objects appear not to have been detected in previous surveys.

\subsection{Radio-Excess Objects}
\label{sec:res_rex}

We select a subsample of the radio-excess objects from the PMN/FSC
sample based on the ratio of their FIR and radio fluxes, $u =
\log~[\Snfir/\Snrad]$ (CAB). Objects on the radio-quiet radio/FIR
correlation have $u = 2.5\pm0.7$. Radio-excess objects were defined by
\citet{yun01} to have radio emission more than five times larger than
the radio/FIR correlation (i.e., star formation contributes $<$ 20\%
of the radio emission from these objects). We adopt this criterion,
which corresponds to values of $u < 1.8$.  Objects that are
traditionally classed as radio-loud have $u \leq 0.0$. We refer to
objects with $0.0 < u < 1.8$ as radio-intermediate objects. These
objects lie between the radio-quiet and radio-loud regions marked in
Fig.\ \ref{fig:lums}. We focus on the nature of the radio-excess
objects in the remainder of the paper.

The radio-excess sample contains 98 PMN/FSC objects and 109 CAB
objects with $u < 1.8$.  These are listed in Table \ref{tab:rad_int}.
Column 1 gives the FSC name. Column 2 indicates the sample from which
the object was selected. Column 3 lists the radio excess parameter,
$u$. Columns 4 and 5 list the object luminosities at 60 \micron\ and
4.8 GHz. Columns 6, 7 and 8 list the radio spectral indices between 3
cm and 6 cm, between 6 cm and 20 cm, and between 6 cm and 36 cm,
respectively.  We use our 3 and 6 cm flux densities obtained with the
\atca. The 20 cm flux densities are from the NVSS.  The 36 cm flux
densities are from the 843 MHz Sydney University Molongolo Sky Survey
\citep[SUMSS]{mau03}.  In all cases, the spectral index is defined as
$S_{\nu} \propto \nu^{\alpha}$. Column 9 lists the projected linear
size of the radio source measured from our \atca\ maps or taken from
the literature.  Column 10 lists the reference from which this value
has been derived.  Column 11 describes the object type.  Column 12
lists the object identifications as given in Column 13 of Table
\ref{tab:sam}.

Identification charts for the PMN/FSC radio-excess objects are
presented in Fig.\ \ref{fig:charts} where we indicate the radio
position, \iras\ $1\sigma$ and $3\sigma$ error ellipses, and our
optical identification. Identification charts for some CAB objects
were given in that paper.

\placetable{tab:rad_int}
\begin{deluxetable}{llrccrrrrcll}
\tabletypesize{\scriptsize}
\rotate
\tablewidth{\textwidth}
\tablewidth{0pt}
\tablecolumns{12}
\tablecaption{Radio-Excess Objects\label{tab:rad_int}}
\tablehead{
\colhead{FSC Name} &
\colhead{Origin} &
\colhead{$u$} &
\colhead{$\log \nu L_\nu$(60 \micron)} &
\colhead{$\log L_\nu$(4.85 GHz)} &
\colhead{$\alpha_{3-6 cm}$} &
\colhead{$\alpha_{6-20 cm}$} &
\colhead{$\alpha_{6-36 cm}$} &
\colhead{Radio Size} &
\colhead{Ref} &
\colhead{Object} &
\colhead{Identification} \\
& & &
\colhead{(\Lsun)} &
\colhead{(\whz)} &
& & &
\colhead{(kpc)} &
&
\colhead{Type} & \\
\colhead{(1)} &
\colhead{(2)} &
\colhead{(3)} &
\colhead{(4)} &
\colhead{(5)} &
\colhead{(6)} &
\colhead{(7)} &
\colhead{(8)} &
\colhead{(9)} &
\colhead{(10)} &
\colhead{(11)} &
\colhead{(12)}
}
\startdata
F00022+1231 & CAB &  -0.181 &  11.770 &  25.773 & \nodata\phn & 1.53\phn\phn&  \nodata\phn&921.0\phn\phn\phn&\phn8& \nodata    & PKS~0002+12     \\
F00123-2356 & PMN &\phs   0.621 &  10.960 &  24.185 & -1.68\phn\phn& -0.42\phn\phn&  \nodata\phn&72.9\phn\phn\phn&\phn1& \nodata    & Abell~14        \\
F00129+1422 & CAB &\phs   0.499 & \nodata & \nodata & \nodata\phn & -0.62\phn\phn&  \nodata\phn&\nodata\phn\phn\phn&\nodata& E          & \nodata         \\
F00183-7111 & PMN &\phs   1.042 &  12.918 &  25.635 & -1.25\phn\phn&  \nodata\phn& -0.78\phn\phn&$<$5.8\phn\phn\phn&\phn1& ULIRG      & IRAS~00182-7112 \\
F00349+0916 & CAB &\phs   1.046 & \nodata & \nodata & \nodata\phn & -0.82\phn\phn&  \nodata\phn&\nodata\phn\phn\phn&\nodata& \nodata    & UGC~00373       \\
F00460+3141 & CAB &\phs   0.721 &  10.219 &  23.372 & \nodata\phn & -0.14\phn\phn&  \nodata\phn&$<$1.3\phn\phn\phn&\phn9& S0         & NGC~262         \\
F00550+3004 & CAB &  -0.829 &\phn   9.674 &  24.376 & \nodata\phn & 0.69\phn\phn&  \nodata\phn&1749.6\phn\phn\phn&12& E, FR~I    & NGC~315         \\
F00567+2647 & CAB &\phs   0.969 &  10.906 &  23.799 & \nodata\phn & -0.37\phn\phn&  \nodata\phn&40.0\phn\phn\phn&12& \nodata    & IC~64           \\
\enddata
\protect\tiny{
\tablenotetext{a}{CAB position is correct, but identification chart is incorrect.} 
\tablenotetext{b}{GPS source discussed by \citet{cra96}.} 
\tablenotetext{c}{Galaxy pair, both radio sources. Tabulated radio fluxes and optical position are for eastern galaxy. $S_\nu$(4.8 GHz) = 17.4 mJy and $S_\nu$(8.6 GHz) = 13.6 mJy and 13$^h$20$^m$04$^s$.7 -17$^\circ$07\arcmin 17\arcsec\ for western galaxy.} 
\tablerefs{
(1) Our ATCA data; (2) \citealt{alt95}; (3) \citealt{ant85}; (4) \citealt{bla92};
(5) \citealt{bri81}; (6) \citealt{bro98}; (7) \citealt{cas99}; (8) \citealt{con95};
(9) \citealt{con88}; (10) \citealt{con91}; (11) \citealt{con82}; (12) \citealt{con91b};
(13) \citealt{cra96}; (14) \citealt{fan86}; (15) \citealt{fan87}; (16) \citealt{fos98};
(17) \citealt{gab00}; (18) \citealt{gho95}; (19) \citealt{hut94}; (20) \citealt{lea97};
(21) \citealt{maz91}; (22) \citealt{mor99a}; (23) \citealt{mor99b}; 
(24) \citealt{ode98}; (25) \citealt{ode88}; (26) \citealt{owe00}; (27) \citealt{ped90};
(28) \citealt{sch01}; (29) \citealt{she97}; (30) \citealt{tad01}; (31) \citealt{the00};
(32) \citealt{ulv86}; (33) \citealt{ulv83}; (34) \citealt{wal87}; (35) \citealt{weh92};
(36) FIRST image via NED.
Published angular sizes were converted to linear sizes using
$H_0 = 50$ \kms~Mpc$^{-1}$ and $q_0 = 0.5$.}
}
\end{deluxetable}

\placefigure{fig:charts}
\begin{figure}
\figurenum{3}
\caption{{\bf Please note: Figures \ref{fig:charts}a -- d are provided
separately in GIF format.} Identification charts for the radio-excess PMN/FSC
objects taken from DSS red images. Crosses mark radio sources (PMN, \atca\, or
NVSS). Solid and dashed ellipses indicate the \iras\ $1\sigma$ and $3\sigma$
positional uncertainties, respectively.  Circles mark the positions of the
optical identifications. Each chart is 3 arcmin on a side with North up and
East to the left. The linear scale at the object corresponding to an angle of
1 arcmin is shown at lower right. The object redshift is listed at lower
left.\label{fig:charts}}
\end{figure}

\clearpage

\stepcounter{figure}

\section{Well-Studied Objects in the Radio-Excess Sample}
\label{sec:kno}

Several well-studied objects are present in the radio-excess sample.
While it is not possible to generalize about the nature of all
radio-excess objects from the properties of a few, their presence does
provide some insight into the types of objects that constitute the
radio-excess population.  The well-studied radio-excess objects are
all AGN of various types including interacting Seyfert galaxies,
CSS/GPS and double-lobed radio galaxies, radio-loud quasars, and
blazars.  Their radio powers range over more than 6 orders of
magnitude, and their radio sizes span the range from kpc-scale objects
smaller than the host galaxy to Mpc-scale extended radio galaxies.
Compact radio sources occur in both Seyfert galaxies, where the radio
source has low power and small radio excess, and in CSS and GPS
sources that have high radio powers and large radio excesses.  The
well-studied objects with extended radio structure all have large
radio-excesses and high radio powers. The FIR luminosities of the
well-studied radio-excess objects range over at least 4 orders of
magnitude.  AGN-dominated ultra-luminous infrared galaxies are present
at high FIR luminosities. Most of these have large radio excesses.
Only a few well-studied FIR-luminous objects have intermediate
radio-excesses. At low FIR luminosities, elliptical galaxies with
compact radio cores are common. Many of these have intermediate
radio-excesses. We now consider each of these objects in detail.

\subsection{Ultra-Luminous Infrared Galaxies}
\label{sec:kno_uli}

IRAS 00182-7112 (F00183-7111) is a high redshift ($z = 0.327$) ULIRG
($\nLnfir = 10^{12.8} \> \Lsun$). It is one of the few well-studied
high FIR luminosity objects having an intermediate radio excess ($u =
1.04$).  It is a powerful radio source ($\Lnrad = 10^{25.6}$ \whz)
that is unresolved in our 1\arcsec\ \atca\ beam ($<$ 5.8 kpc) at 8.6
GHz. \citet{nor88} detected a compact radio core at 1.66 GHz using the
Parkes-Tidbinbilla Interferometer. They measured a flux density of 250
mJy with fringe spacing $\sim$ 0.1\arcsec, which corresponds to 578
pc. The flux density expected for the source at 1.66 GHz, derived from
the observed ATCA flux density at 5 GHz and SUMSS flux density at 843
MHz, is 249 mJy. This strongly suggests that the radio source is
compact, with total extent $<$ 578 pc. There is evidence for a weak or
obscured AGN from the classification of the optical spectrum as a
LINER \citep{arm89}. Strong mid-infrared continuum and relatively weak
polycyclic aromatic hydrocarbon (7.7 \micron; PAH) emission indicate
the presence of an obscured AGN \citep{tra01}.

\subsection{Seyfert Galaxies}
\label{sec:kno_sey}

The Superantennae (F19254-7245) is an infrared luminous interacting
pair of galaxies ($\nLnfir = 10^{12.05} \> \Lsun$). It has a small
radio excess ($u = 1.76$) and moderate radio power ($\Lnrad =
10^{24.2}$ \whz).  The small radio excess is only just sufficient to
distinguish it from radio-quiet objects. The two distinct nuclei have
been classified as a Seyfert 2 (southern galaxy) and a starburst/LINER
(northern galaxy) based on optical spectroscopy \citep{mir91}.  The
Seyfert galaxy is the dominant radio source (\atca\ data) and is
likely to be the FIR source (Mirabel et al. 1991).  The radio source
is smaller than our \atca\ beam at 8.6 GHz so must be $<$ 500 pc in
extent.

Mrk 463 (F13536+1836) contains two nuclei that are separated by $\sim$
5.9 kpc and embedded in a common envelope \citep{maz93}.  Both nuclei
have Seyfert~2 optical spectra \citep{shu81} though the eastern
nucleus, Mrk 463E, has a Seyfert 1 nucleus observed in polarized light
(Veilleux, Sanders, \& Kim 1997).  Both nuclei are radio sources. Mrk
463E has a luminous steep-spectrum core and weak radio lobes with
extent $\sim$ 27 kpc \citep{maz91}. \citet{maz91} suggest Mrk 463E may
be a transition object between the smaller, weaker radio sources seen
in Seyfert galaxies and extended powerful radio galaxies. At high
resolution, the core is found to be a compact double with extent
$\sim$ 1.8 kpc \citep{the00}. Mrk 463W has a weaker radio source with
a flatter spectrum \citep{maz91}. Overall, Mrk 463 appears in our
sample as a moderately powerful radio source ($\Lnrad = 10^{24.1}$
\whz) with high FIR luminosity ($\nLnfir = 10^{11.5} \> \Lsun$) and an
intermediate radio excess ($u = 1.30$).

NGC 7212 (F22045+0959) is a Seyfert 2 galaxy in an interacting triple
\citep{was81}. It has a small radio excess of $u = 1.80$ on the
boundary with radio-quiet objects. The radio source is a compact
double with a separation of $\sim$ 520 pc \citep{fal98} and moderate
radio power ($\Lnrad = 10^{23.2}$ \whz). NGC 7212 is luminous
in the FIR, having $\nLnfir = 10^{11.1} \> \Lsun$.

NGC 2110 (F05497-0728) is a nearby X-ray luminous Seyfert~2 galaxy
with an intermediate radio excess of $u = 1.40$.  NGC 2110 has a small
radio source with a low radio power of $\Lnrad = 10^{22.7}$ \whz. The
radio source consists of a compact flat-spectrum core and symmetric
jets with a total extent of $\sim$ 590 pc \citep{mun00}.  The optical
host is an elliptical galaxy with an equatorial dust lane, which is
indicative of a previous merger \citep{col01}, and a moderate FIR
luminosity of $\nLnfir = 10^{10.2} \> \Lsun$.

NGC 5506 (F14106-0258) is also a nearby X-ray luminous Seyfert~2
galaxy with a small radio excess of $u = 1.68$. The radio source has
very low power ($\Lnrad = 10^{22.4}$ \whz) and is small with an
unresolved core and diffuse emission extending over $\sim$ 500 pc
\citep{the00}.  The host is an Sa galaxy with a moderate FIR
luminosity of $\nLnfir = 10^{10.2} \> \Lsun$.

IC 5063 (F20482-5715) has properties in common with both Seyfert
galaxies and radio galaxies \citep{col91}. It is an X-ray bright
Seyfert 2 galaxy, with intermediate radio excess ($u = 1.16 $) and low
radio power ($\Lnrad = 10^{23.5}$ \whz).  The radio emission arises
from a compact double-lobed source 1.3 kpc in extent \citep{mor99b}.
The host galaxy is a nearby elliptical with dust lanes, so it has
probably undergone a recent merger. The galaxy has a moderate FIR
luminosity of $\nLnfir = 10^{10.6} \> \Lsun$.

\subsection{LINERs}
\label{sec:kno_lin}

M104 (F12374-1120, Sombrero Galaxy, NGC 4594) is a well-known edge-on
disk galaxy with a large bulge. It has an intermediate radio excess of
$u = 1.50$ and very low radio power of $\Lnrad = 10^{21.9}$ \whz. It
also has a low FIR luminosity of $\nLnfir = 10^{9.4} \> \Lsun$. The
optical emission-line spectrum is classified as a LINER and the
optical continuum is dominated by an old stellar population
\citep{kin93}. Our \atca\ observations show the GHz spectral slope to
be slightly inverted.  The radio source has a compact core less than
31 pc in extent \citep{the00}, and extended radio emission that is
weak relative to the central source and may be due to star formation
or indirectly related to the AGN \citep{baj88}.

\subsection{Ellipticals With Radio Cores}
\label{sec:kno_ell}

Several radio-excess galaxies belong to the sample of elliptical
galaxies with radio cores studied by \citet{sle94} and mentioned in
\S\ref{sec:res_rad}: NGC 612 (F01317-3644), NGC 1052 (F02386-0828),
NGC 2110 (F05497-0728, discussed above), NGC 2911 (09311+1022), IC
5063 (F20482-5715, discussed above), NGC 7213 (F22061-4724), and IC
1459 (F22544-3643).  The radio cores in elliptical galaxies were
typically found by these authors to be extremely compact (unresolved
on scales of a few parsecs) and to have a large range of radio powers
($\Lnrad = 10^{21}-10^{26}$ \whz). Their high-frequency spectral
indices are typically inverted or flat (${<\alpha>} \approx 0.3$,
$S_{\nu} \propto \nu^{\alpha}$), due to absorption consistent with SSA
or FFA.  These objects have a range of radio power and radio excess,
but low FIR luminosities ($\nLnfir < 10^{11} \> \Lsun$).

\subsection{Radio Galaxies}
\label{sec:kno_rad}

3C~195 (F08064-1018, PKS 0806-10) is a high-power radio galaxy
($\Lnrad = 10^{25.9}$ \whz) with a large radio excess ($u = -0.51$)
that justifies its radio-loud classification. It has an FR~II
morphology \citep{fan74} with a total extent of $\sim$ 367 kpc
\citep{mor93}. The optical host is an asymmetric elliptical galaxy
that appears to be interacting with a small less-luminous companion
$\sim$ 11 kpc (projected linear separation) from the nucleus. 3C~195
is a luminous FIR galaxy with $\nLnfir = 10^{11.5} \> \Lsun$.

3C~327 (F15599+0206, PKS 1559+02) is also a powerful FR~II radio
galaxy. It has $\Lnrad = 10^{26.0} \> \whz$ and a large radio excess of
$u = -0.67$.  The asymmetric radio lobes extend to $\sim$ 768 kpc
\citep{lea97}. The optical host is a flattened elliptical with a
Seyfert 2 nucleus.  It is also a luminous FIR source with $\nLnfir =
10^{11.5} \> \Lsun$.

PKS 0634-20 (F06343-2032) is a radio galaxy with $\Lnrad = 10^{25.7}$
\whz\ and a very large radio excess ($u = -0.88$). It is an FR~II with
a total extent of nearly 1.3 Mpc (CAB).  It has a strong
high-ionization optical spectrum and an apparently normal elliptical
host. The FIR luminosity is moderate with $\nLnfir = 10^{10.9} \>
\Lsun$.

NGC 315 (F00550+3004) is a radio galaxy with a very large radio excess
of $u = -0.83$, but with only moderate radio power ($\Lnrad =
10^{24.4}$ \whz). It is a large FR~I galaxy with a total extent of $>$
1.7 Mpc \citep{con91b}. Evidence for superluminal motion has been
detected in the radio source \citep{xu00}.  The host is an elliptical
galaxy with an equatorial dust disk and an unresolved optical nucleus
\citep{cap00}. It has a low FIR luminosity of $\nLnfir = 10^{9.7} \>
\Lsun$.

3C~120 (F04305+0514) is a powerful radio galaxy ($\Lnrad = 10^{25.4}$
\whz) with a large radio excess of $u = -0.43$. It has FR~I morphology
on large scales with a total extent of $>$ 760 kpc (Walker et
al. 1987).  Superluminal motion has been detected in the small-scale
radio structure and an optical and radio jet is observed on arcsecond
scales. This object is highly variable at all wavelengths (radio --
X-ray) and may be a blazar. The optical nucleus has a Seyfert 1
spectrum and the host galaxy may be disturbed. The FIR luminosity is
moderate at $\nLnfir = 10^{10.9} \> \Lsun$.

NGC 1275 (F03164+4119, Perseus A) is the central cD galaxy in the
Perseus cluster. It has a large radio excess ($u = -0.77$) and is
luminous in both the radio and FIR with $\Lnrad = 10^{25.8}$ \whz\ and
$\nLnfir = 10^{11.1} \> \Lsun$. The radio source has an asymmetric
FR~I morphology on kilo-parsec scales and complex morphology on
smaller scales.  It has a spiral companion at a separation of $\sim$
3000 \kms\ that may be infalling. The optical spectrum is classified
as a Sy~1.5 but \citet{ver78} has suggested the reclassification of
this galaxy to a BL Lac due to its optical polarization and
variability.

PKS 1549-79 (F15494-7905) is a powerful, compact, flat-spectrum radio
source with \Lnrad = 10$^{26.6}$ \whz\ and a large radio excess of $u
= -0.56$. It is a Narrow Line Radio Galaxy (NLRG) with a
core-dominated core-jet morphology of total extent $\sim$ 512 pc
\citep{tad01}. It is FIR luminous with $\nLnfir = 10^{12.08} \>
\Lsun$.  \citet{tad01} argue that this is a young radio source that is
intrinsically compact and embedded in a dense host galaxy interstellar
medium, similar to CSS/GPS sources. They argue that the axis of
radio emission is close to the line of sight, resulting in the
observed flat radio spectrum, but that the quasar nucleus is highly
obscured at optical wavelengths. The host galaxy is not well studied,
but appears on Schmidt plates to be an elliptical with disturbed outer
morphology \citep{jau89}.

M 87 (F12282+1240, NGC 4486, 3C~274) is a giant elliptical galaxy at
the center of the Virgo cluster. It has the largest radio excess in
the sample with $u = -2.25$. The radio source has moderate power with
$\Lnrad = 10^{24.6}$ \whz\ and extended FR~I morphology
\citep{mar99}. The galaxy has a bright optical synchrotron jet and a
LINER spectrum \citep{dop97}. The FIR emission has low luminosity with
$\nLnfir = 10^{8.5} \> \Lsun$.

\subsection{CSS/GPS Sources}
\label{sec:kno_css}

PKS 1345+12 (F13451+1232, 4C 12.50) is a nearby radio-loud GPS galaxy
with a radio excess of $u = -0.21$. The radio source is powerful
($\Lnrad = 10^{26.3}$ \whz).  The FIR emission is very luminous at
$\nLnfir = 10^{12.2} \> \Lsun$. The host galaxy is one of an
interacting pair.  The optical nuclei are separated by 4.3 kpc and are
embedded in an asymmetric common envelope 43 kpc across \citep{hec86}.
The south-eastern elliptical galaxy was previously thought to be the
radio source \citep{gil86}, but Hubble Space Telescope observations
combined with high-resolution radio data have shown the radio emission
to be associated with the north-western nucleus \citep{eva99}. This
galaxy has a Seyfert 2 spectrum with a hidden broad-line region
observed in the near-infrared \citep{vei97}.  This object provides
support for the scenario in which galaxy mergers produce high FIR
luminosities and funnel gas to the nucleus to fuel an AGN
\citep{san88a}.  A high mass of molecular gas ($>$ 10$^{10}$ \Msun)
has been inferred from observations of \ion{H}{1} absorption
\citep{mir89} and CO emission \citep{eva99}.  The CO emission is
compact, suggesting the gas is concentrated around the nuclear region
and provides fuel for the AGN \citep{eva99}.  The continuum radio
source is a two-sided jet with total extent $\sim$ 290 pc
\citep{lis03}. It is likely that the radio source is young, assuming
that the jets are propagating at velocities similar to other compact
symmetric objects \citep{lis03}.  Evans et al. (1999, 2002) argue that
the molecular gas is the source of fuel for the radio emission, as CO
is only detected in the more radio-luminous nucleus in several
interacting galaxy pairs. It is not clear, however, if the high
molecular mass triggers or simply fuels the radio source. The detailed
VLBI study of PKS 1345+12 by \citet{lis03} presents evidence that the
radio jet is precessing, which suggests that the radio jet may have
been triggered by the merger of two black holes, producing a spinning
black hole.

3C~459 (F23140+0348) is a CSS galaxy with a large radio excess of $u =
-0.35$. It is luminous at radio and FIR wavelengths with $\Lnrad =
10^{26.4}$ \whz\ and $\nLnfir = 10^{12.2} \> \Lsun$.  The radio source
has a core and two lobes separated by 40 kpc. The host galaxy shows a
post-starburst optical spectrum with blue continuum, young stellar
absorption lines \citep{mil81}, and nebular emission lines that
\citet{zhe99} classified as a LINER. The host is an elliptical with
disturbed morphology in the outer parts, possibly the remnant of
multiple mergers \citep{zhe99}.

3C~48 (F01348+3254) is an extremely radio-loud ($u = -0.89$) CSS
quasar.  It is very luminous in both the radio and FIR with $\Lnrad =
10^{27.4}$ \whz\ and $\nLnfir = 10^{12.7} \> \Lsun$.  The radio source
is small and consists of a flat-spectrum core and a bright
steep-spectrum lobe \citep{wil91} with a total extent of $\sim$ 8 kpc
\citep{spe89}. The host galaxy is overwhelmed at optical and
near-infrared wavelengths by the quasar nucleus, but subtraction of
the central source reveals disturbed outer isophotes indicative of a
merger \citep{sco00, sto91}.  The optical spectrum is unusual for a
quasar in that it shows strong Balmer absorption lines that indicate
the presence of a dominant young stellar population \citep{bor84}.

3C~298 (F14165+0642) is also an extremely radio-loud ($u = -0.86$) CSS
quasar.  It is extremely luminous at radio and FIR wavelengths with
$\Lnrad = 10^{27.9}$ \whz\ and $\nLnfir = 10^{13.2} \> \Lsun$. The
radio source has a compact triple morphology and the host galaxy may
be disturbed.

\subsection{Quasars}
\label{sec:kno_qua}

There are a number of known radio-loud quasars in the radio-excess
sample: PKS 0104-408 (F01044-4050), Q0521-365 (F05212-3630), Q0558-504
(F05585-5026), Q0637-752 (F06374-7513), Q1244-255 (F12440-2531),
3C~345 (F16413+3954), PKS 2313-477 (F23135-4745), and Q2349-014
(F23493-0126). These objects all have high radio powers ($\Lnrad
\gtrsim 10^{25}$ \whz) and FIR luminosities $\nLnfir > 10^{10.7} \>
\Lsun$. All but one of these quasars have large radio excesses with $u
< -0.2$; Q0558-504 is the exception.

Q0558-504 (F05585-5026) is the only previously-known quasar with an
intermediate radio excess. This object has $u = 0.39$. It is a
luminous FIR and radio source with $\Lnrad = 10^{25.0}$ \whz\ and
$\nLnfir = 10^{11.3} \> \Lsun$.  It has a Narrow Line Seyfert 1
optical spectrum with bright and variable X-ray emission
\citep{bal01}.

\subsection{Blazars}
\label{sec:kno_bla}

Blazars are common in the radio-excess sample; PKS~0235+164
(F02358+1623), PKS~0338-214 (F03384-2129), PKS~0420-014 (F04207-0127),
PKS~0537-441 (F05373-4406), PKS~0735+17 (F07352+1749), PKS~0754+100
(F07543+1004), PKS~0829+046 (F08291+0439), OJ~287 (F08519+2017),
PKS~1144-379 (F11445-3755), 3C~273 (F12265+0219), 3C~279
(F12535-0530), B2~1308+32 (F13080+3237), OQ~208 (F14047+2841), OQ~530
(F14180+5437), B2~1732+389 (F17326+3859), Q2005-489 (F20057-4858),
BL~Lac (F22006+4202), and 3C~446 (F22231-0512). These objects are BL
Lacs, optically variable, flat radio spectrum quasars, or ``transition
objects'' between traditional BL Lacs and (strong emission line)
quasars.  Many of the blazars have extremely high radio powers
($\Lnrad > 10^{27}$ \whz) and FIR luminosities ($\nLnfir > 10^{13} \>
\Lsun$) and are at relatively large redshifts ($z \gtrsim
0.9$). Blazars also occur in the sample at lower redshifts and powers,
as low as $z \sim 0.05$ and $\Lnrad \sim 10^{25}$ \whz. All the known
blazars in the radio-excess sample have large radio excesses ($u <
-0.2$), several with extreme values of $u \sim -1.0$. CAB comment that
the blazars in their full sample all have $u < -0.15$ and spectral
indices between 1.4 and 4.8 GHz of $<$ 0.5. This makes them flat
spectrum objects with large radio excesses.

\section{Discussion}
\label{sec:dis}

\subsection{Radio Properties}
\label{sec:dis_rad}

The radio-excess sample includes radio sources with a large range of
physical sizes from $<$~1~kpc to $>$~1~Mpc.  Angular sizes or upper
limits were measured from our \atca\ observations for 51 radio-excess
objects and physical sizes determined using the known redshifts. Sizes
for a further 116 objects were obtained from the literature.  Sizes
for the radio sources and references are listed in Table
\ref{tab:rad_int}.  Of the 166 objects for which we have radio sizes,
102 are $\le$ 15 kpc in extent, implying that the radio source is
embedded within the host galaxy in many cases. Of these, 7 are known
blazars or BL Lac objects, leaving 95 compact non-blazar objects
(57\%\ of objects with measured sizes). Figures \ref{fig:atca5} and
\ref{fig:atca8} show maps of the 17 sources observed with the \atca\
that were resolved at 4.8 GHz and 8.6 GHz, respectively. The peak
radio fluxes and contour levels are listed in Table \ref{tab:con}. The
radio contours are overlaid on DSS optical images. One of these
objects (F13174-1651) is a pair of galaxies separated by 14 kpc. The
eastern galaxy is resolved with a size of 5.5 kpc, the western galaxy
is unresolved ($<$ 1 kpc).  Three of the radio sources (F03079-3031,
F08064-1018 and F20203-5733) are large radio galaxies with physical
sizes of several hundred kpc. Only the core of F03079-3031 is shown in
Fig.\ref{fig:atca5} as the lobes were not detected. Another object
(F22073-4044, redshift unknown) has lobes separated by 41\arcsec\ and
may also be a large radio galaxy.  Five of the resolved sources have
physical sizes less than 15 kpc; all except one of these (F05212-3630)
appear from the DSS images to be similar in size to the optical
sources. The remaining 7 objects have sizes of a few tens of kpc and
are generally larger than the optical source.

\placefigure{fig:atca5}
\begin{figure}
\caption{{\bf Please note: Figure \ref{fig:atca5}\ is provided
separately in GIF format.} \atca\ maps of radio-excess objects that are resolved at 4.8
GHz.\label{fig:atca5}}
\end{figure}

\placefigure{fig:atca8}
\begin{figure}
\caption{{\bf Please note: Figure \ref{fig:atca8}\ is provided
separately in GIF format.} \atca\ maps of radio-excess objects that are resolved at 8.6
GHz.\label{fig:atca8}}
\end{figure}
\placetable{tab:con}
\begin{deluxetable}{llcl}
\tabletypesize{\scriptsize}
\tablewidth{0pt}
\tablecolumns{4}
\tablecaption{Contour levels for \atca\ maps\label{tab:con}}
\tablehead{
\colhead{FSC Name} &
\colhead{$\nu$} &
\colhead{Peak} &
\colhead{Contour Levels} \\
&
\colhead{(GHz)} &
\colhead{(mJy/beam)} &
\colhead{(\% of peak)} \\
\colhead{(1)} &
\colhead{(2)} &
\colhead{(3)} &
\colhead{(4)} \\
}
\startdata
F00123-2356 & \phn 4.79 &    \phn\phn\phn 3.1 & 20, 30, 40, 50, 70, 90 \\ 
F01317-3644 & \phn 4.79 &   \phn\phn 31.4 & 7, 10, 30, 50, 70, 90 \\ 
F03079-3031 & \phn 4.79 &   \phn\phn 25.5 & 5, 10, 20, 30, 50, 70, 80, 90 \\ 
F03265-2852 & \phn 4.79 &  \phn 265.2 & 1, 2, 5, 8, 10, 30, 50, 70, 80, 90 \\ 
F04367-2726 & \phn 8.64 &    \phn\phn\phn 8.4 & 10, 30, 50, 70, 80, 90 \\ 
F05212-3630 & \phn 4.79 & 2292.0 & 2, 5, 10, 30, 50, 70, 80, 90 \\ 
F08064-1018 & \phn 4.79 & \phn 195.7 & 6, 10, 20, 30, 50, 70, 90 \\ 
F10227-8251 & \phn 4.79 & \phn\phn 11.6 & 10, 20, 30, 50, 70, 90 \\ 
F13174-1651 & \phn 4.79 & \phn\phn  17.2 & 4, 10, 20, 30, 50, 70, 90 \\ 
F20203-5733 & \phn 4.79 & \phn 243.4 & 5, 10, 20, 30, 50, 70, 90 \\ 
F21529-6955 & \phn 4.79 & \phn 809.0 & 6, 10, 14, 20, 30, 50, 70, 90 \\ 
F22073-4044 & \phn 4.79 & \phn 111.6 & 15, 25, 50, 70, 90 \\ 
F22521-3929 & \phn 4.79 & \phn 171.9 & 1, 2, 5, 10, 20, 30, 50, 70, 90 \\ 
F22537-6512 & \phn 8.64 &  \phn\phn 50.2 & 5, 10, 20, 30, 50, 70, 80, 90 \\ 
F23135-4745 & \phn 8.64 & \phn 142.8 & 1, 2, 5, 10, 20, 30, 50, 70, 90 \\ 
\enddata
\end{deluxetable}

\clearpage

\begin{figure}[ht]
\plottwo{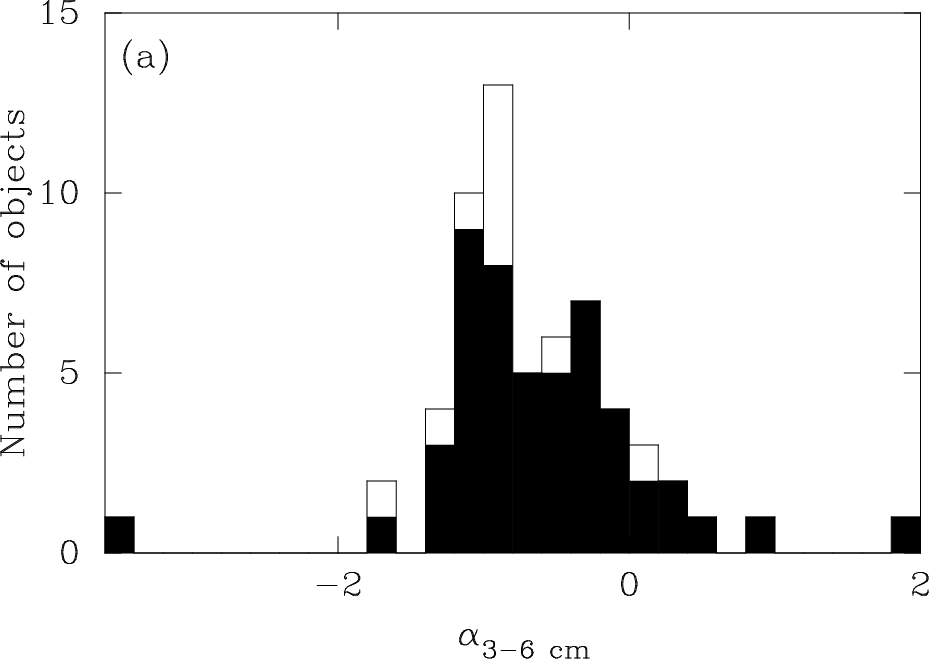}{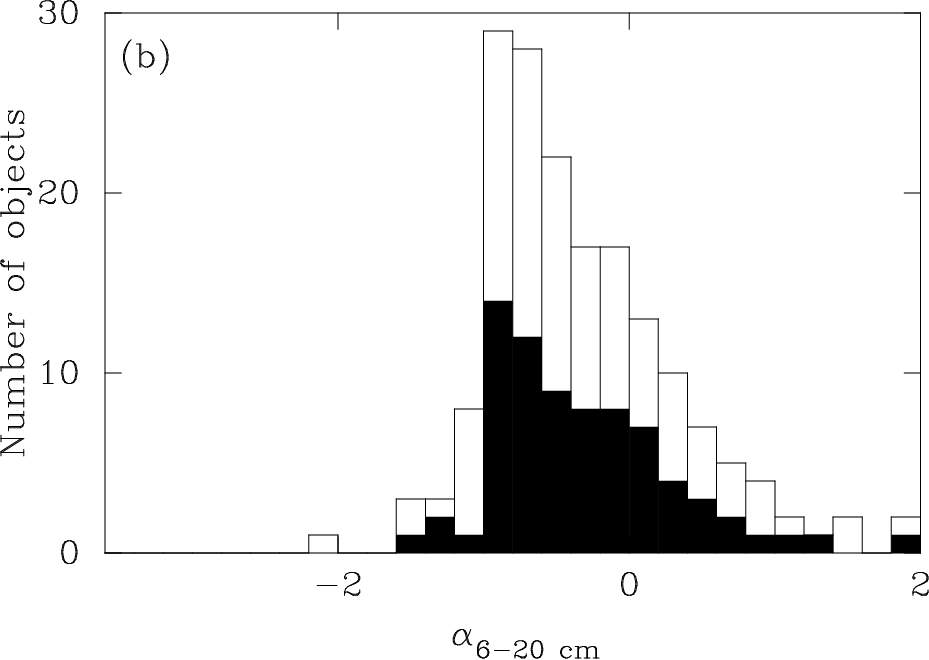}
\caption{(a) Histogram of the radio spectral index, $\alpha_{3-6~cm}$
($S_\nu \propto \nu^\alpha$), between 3 and 6 cm for the radio-excess
objects observed with the \atca. (b) Histogram of the radio spectral
index, $\alpha_{6-20~cm}$, between 6 and 20 cm for the radio-excess
objects with NVSS 20 cm flux densities. Objects with radio source
extent $\le$ 15 kpc, excluding known blazars, are shown filled,
while the unfilled histogram shows the whole sample.\label{fig:alphist}}
\end{figure}

Figure \ref{fig:alphist}(a) shows the distribution of high frequency
spectral indices for the radio-excess sources detected at 3 cm and 6
cm (8.6 GHz and 4.8 GHz, respectively) with the \atca.  Figure
\ref{fig:alphist}(b) shows the distribution of 6--20 cm spectral
indices for those sources in the radio-excess sample with NVSS 20 cm
flux measurements.  To facilitate a comparison with CSS/GPS sources,
non-blazar radio sources $\le$ 15 kpc in extent are indicated
separately in Figure \ref{fig:alphist}.  Most 3 -- 6 cm spectral
indices are in the range -1.5 -- +1.0, with a modes of -0.9 for the
whole radio-excess sample and -1.1 for the compact ($<$ 15 kpc)
objects. This is similar to other types of radio sources, including
CSS/GPS sources \citep{sta98, fan01}.  A large fraction of the compact
objects (28 of 47 sources observed) have steep spectral indices
($\alpha_{3-6~cm} < -0.5$).  The 6--20 cm spectral indices cover a
similar range, with a mode of -0.9 for both the compact objects and
for the whole radio-excess sample.  These steep spectral indices are
similar to those observed in Seyfert radio sources, which have on
average $\alpha_{6-20~cm} \sim -0.7$ \citep{ede87a}.  The spectral
index of F05265-4720 is highly inverted ($\alpha_{3-6~cm} =
1.83$). This suggests either strong synchrotron self-absorption in a
fairly homogeneous medium, or free-free absorption in a dense ionized
medium. At the other extreme, F04023-1638 shows a very steep spectral
index of -3.43. The spectral energy distribution of this source
remains steep to 20 cm but with a shallower slope of -1.37. The cause
of this extremely steep spectrum is not clear.

\placefigure{fig:alphist}

The compact radio-excess objects with steep spectral indices appear to
be similar to CSS or GPS sources. This is consistent with the presence
of known CSS/GPS sources in the sample (\S\ref{sec:kno_css}). CSS/GPS
sources are powerful, optically-thin radio emitters that are also
compact. They have been modeled using the same twin-jet models that
apply to extended radio galaxies (e.g., Carvalho 1985; Begelman 1996).
These models assume that twin jets of highly energetic and collimated
relativistic plasma are expelled by the central engine into the
surrounding interstellar medium of the host galaxy.  A cocoon of
shocked ISM material expands around the jets as the lobes develop. Hot
spots (i.e., regions of high radio brightness) occur at the ends of
the cocoon where the jet interacts directly with the ISM and where the
cocoon expands fastest. The radio emission from the lobes and hot
spots is optically thin at high frequencies, and dominates the total
radio emission from the source.  This results in the observed steep
high frequency radio spectrum.  The small size of CSS and GPS sources
is argued to be due either to youth or confinement: they may be small
because they are very young ($\lesssim 10^5$ yr) and will evolve into
larger radio galaxies, or they may be small because they are confined
by a dense surrounding medium that does not allow the radio lobes to
expand.

Recent evidence has favored the youth model. The measurement of
hotspot advance speeds of $\sim$ 0.2$c$ \citep{ows98, ows99} show that
at least some of the compact sources are expanding, and so are not
permanently confined by the surrounding medium. Plausible evolutionary
models linking CSS/GPS sources with extended radio galaxies (FR~I/II)
also provide support for the youth model \citep{car85, fan95,
rea96}. The possibility that some CSS/GPS sources are confined by the
host ISM has been investigated by numerical simulations and analytical
models. Numerical simulations of a jet propagating in an ambient
medium \citep{you93} suggest that a typical CSS radio jet could be
onfined to a size $<$ 10 kpc for $\sim$ 10$^7$ years by an average
ISM density of $\sim$ 1 -- 10 cm$^{-3}$. This implies total gas masses
in the host of $\sim$ 10$^{11 - 12}$ \Msun. Similar results were
obtained for a uniform density medium and a 2-phase medium consisting
of dense clouds in a diffuse surrounding medium, with the same average
density.  A simple analytical model of jets interacting with dense
clouds in the ISM \citep{car94, car98} has been used to show that some
GPS sources could be confined to sizes less than a few hundred parsecs
for 10$^7$ years by a cloud with central density $\sim$10$^{3-4}$
cm$^{-3}$, assuming a total mass for the cloud of 10$^{9-10}$
\Msun. Gas masses of $\sim$ 10$^{10}$ \Msun\ are known in some CSS/GPS
sources; a mass of $\sim 3 \times 10^{10}$ \Msun\ was inferred for
PKS~B1718-649 from \ion{H}{1} absorption measurements \citep{ver95}
and a mass of $\sim 7 \times 10^{10}$ \Msun\ was derived from CO line
emission for the source PKS~1345+12 (F13451+1232, present in our
sample) \citep{eva99}.  However, a recent study of 49 CSS/GPS sources
found that the best-fit model to the observed \ion{H}{1} column
densities implied a gas mass of $\sim$ 10$^8$ \Msun\ within a radius
of 10 kpc \citep{pih03}. This suggests that masses of 10$^{10}$ \Msun\
are not typical of CSS/GPS sources, so the ambient medium is not
likely to be dense enough to confine the jet in most cases.
Nevertheless, such high gas masses are typical in the central few
hundred parsecs of luminous FIR galaxies \citep{mir88, san96}. The
FIR-luminous radio-excess objects could therefore have gas masses in
the range 10$^{9-10}$ \Msun. The masses and central densities of the
host galaxies may be sufficient to confine the radio
source. Measurements of CO line emission at millimeter wavelengths and
\ion{H}{1} (21 cm) absorption are needed to confirm this.

The radio powers of our compact steep-spectrum radio-excess objects
are generally lower than well-known CSS and GPS sources.  The objects
in our radio-excess sample have radio powers as low as $10^{23}$ \whz,
with a median of $\sim 10^{24.5}$ \whz\ at 5 GHz.  The median radio
power of CSS and GPS sources at 5 GHz is $\sim 10^{27.6}$ \whz\
\citep{ode98}.  This may be contrasted with the radio sources observed
in Seyfert galaxies. These are also subgalactic (usually a few hundred
parsecs; e.g., \citet{sch01, ulv01}) and generally have optically thin
spectra, but have a median power at 5 GHz of $10^{22}$ \whz\
\citep{ede87a}.  The median redshift of the radio-excess sample is
0.063. This is significantly lower than the median redshift of CSS and
GPS samples, which is $z =$ 0.91, and considerably higher than the
median redshift of the Seyfert galaxy sample, 0.0196. Thus a large
part of the difference in radio powers may be attributable to the
correlation between power and redshift for flux-limited
samples. However, the radio jets observed in some Seyfert galaxies are
likely to be intrinsically weaker than those in CSS/GPS sources and
are subrelativistic \citep{bic02}. It is possible that the
radio-excess objects with intermediate radio powers have jet energy
fluxes that fall in between those of Seyfert and CSS/GPS radio
sources. Jet energy fluxes have been estimated from the luminosity of
the \oiii\ emission line, based on shock-excitation models for the
optical emission \citep{bic98}. A comparison of the \oiii\
luminosities of the radio-excess objects with those of Seyferts and
CSS/GPS sources would reveal whether the radio-excess objects have jet
energy fluxes intermediate between Seyferts and CSS/GPS sources.

It has been suggested that CSS/GPS sources may be precursors to FR~I
and II radio galaxies (e.g., Phillips \& Mutel 1982; Carvalho 1985;
O'Dea \& Baum 1997). If CSS sources maintain a constant jet energy
flux as their lobes expand into a uniform medium with radially
declining density, the radio power will decrease due to the decrease
in lobe pressure as the length increases \citep{beg96}.  If powerful
CSS and GPS sources are the precursors of FR~I and II radio galaxies,
the relative numbers of CSS/GPS and extended radio sources demand that
they must decrease in luminosity by factors of 10--100 as they expand
(e.g., O'Dea \& Baum 1997).  More recently, however, it has been
suggested that GPS sources ($<$ 1 kpc in size) may increase in
luminosity as they increase in size if they expand into a constant
density medium \citep{sne00b}. Clearly the circumnuclear environment
of a young radio galaxy is complex and may well be clumpy on small
scales, so these idealized models are only indicative. By fitting
models to observed linear sizes and hot-spot advance speeds of GPS
sources, \citet{per02} conclude that GPS sources increase in
luminosity with linear size up to 1 kpc, lending support to the
suggestion.  A source $\sim$ 10 pc in size might be expected to
increase in luminosity by a factor of $\sim$ 20 as it expands to 1
kpc, assuming that the jet energy flux is constant and that the
density of the surrounding medium is constant with radius.  A
radio-intermediate object with $L_{radio} = 10^{24}$ \whz\ and size
$\sim$ 10 pc (in the ``GPS phase'') could increase to $L_{radio} = 2
\times 10^{25}$ \whz\ as it expanded to 1 kpc. It would then appear to
be a radio-loud object (see Figure \ref{fig:lums}). As it expanded
beyond 1 kpc, it would be expected to decrease in radio power again,
and would possibly evolve into an FR~I or FR~II radio galaxy.

A turnover in the spectra of CSS and GPS sources is seen at
frequencies $\lesssim 15$ GHz \citep{ode97} where the source becomes
optically thick due to either synchrotron self-absorption or free-free
absorption. The observed relationship between the turnover frequency
and emission region size of CSS/GPS sources can be explained by either
absorption model (\citet{ode97} and \citet{bic97}, respectively). At
frequencies around the turnover, the source appears to have a flat
radio spectrum and below this frequency the spectrum is inverted.
Compact objects in the radio-excess sample with flat or inverted
spectral indices at centimeter wavelengths may be young radio sources
observed near or below their spectral peak. These would be sources
with turnover frequencies $\gtrsim$ 1 GHz. The observed relation
between turnover frequency and source size \citep{ode97} would then
imply sizes $\lesssim$ 1 kpc. This is credible because more compact
objects ($\le$ 15 kpc, excluding known blazars) show $\alpha_{6-20cm}$
inverted (26\% of those objects with measured $\alpha_{6-20cm}$ have
$\alpha_{6-20cm} > 0$) than $\alpha_{3-6cm}$ (14\% of those objects
with measured $\alpha_{3-6cm}$ have $\alpha_{3-6cm} > 0$). Fewer
objects show steep slopes between 6 cm and 20 cm than between 3 cm and
6 cm (51\% with $\alpha_{6-20cm} >$ 0.5 compared with 60\% with
$\alpha_{3-6cm} >$ 0.5).  These distributions of spectral slopes at
low and high frequencies are consistent with the presence of GPS
sources in the radio-excess population, which turn over at $\sim$ 1
GHz.

Alternatively, the radio-excess objects with flat radio spectra may be
core-dominated sources.  The degree to which the emission is dominated
by the optically thin lobes or the optically thick core is often used
to infer whether the radio jet is aligned near to the line of sight.
Generally, the closer the radio jet is to the line of sight the more
core-dominated is the source and the flatter is its radio spectrum.
Flat spectrum compact radio sources are usually associated with
radio-loud quasars or BL Lac objects.  The radio-FIR emission from
blazars is believed to be synchrotron emission from relativistic
plasma that is highly collimated and orientated close to the line of
sight. The isotropic radio power of these objects is likely to
over-estimate the true power because of relativistic beaming.  In the
absence of further information on the spectral energy distribution of
the flat-spectrum, compact radio-excess sources, it may be possible to
distinguish the blazars from CSS/GPS-like sources by measuring the
polarization and rotation measure. Blazars are characterized by high
levels of radio polarization ($\sim$ 2.5\%; Iler et al. 1997)
independent of observation frequency \citep{sai88}, while CSS/GPS
sources show weak polarization ($\sim$ 0.2\% for GPS sources and up to
3\% for CSS sources at 5 GHz) which varies with frequency, consistent
with large observed rotation measures \citep{ode98}.

Most objects were not observed to vary in radio flux. Twenty-six
objects were measured with the \atca\ on at least two occassions and
were not found to vary significantly. This is similar to GPS sources,
which are the least variable of radio sources \citep{ode98}. Two of
the radio-intermediate objects observed with the \atca\ were observed
to vary in flux density. F01264-5705, was found to increase in flux
density by a factor of 1.85 at 8.6~GHz between observations separated
by 18 months. F01264-5705 is a compact radio source with a highly
inverted radio spectrum.  F11445-3755 was observed to increase in flux
density by factors of 1.2 at 4.8 GHz and 1.4 at 8.6 GHz within one
month. This object is a known blazar (PKS 1144-379).  However, the
general lack of variability is more consistent with the radio-excess
objects being similar to CSS/GPS sources, rather than blazars.

\subsection{Far-Infrared Properties}
\label{sec:dis_fir}

\begin{figure}[ht]
\plottwo{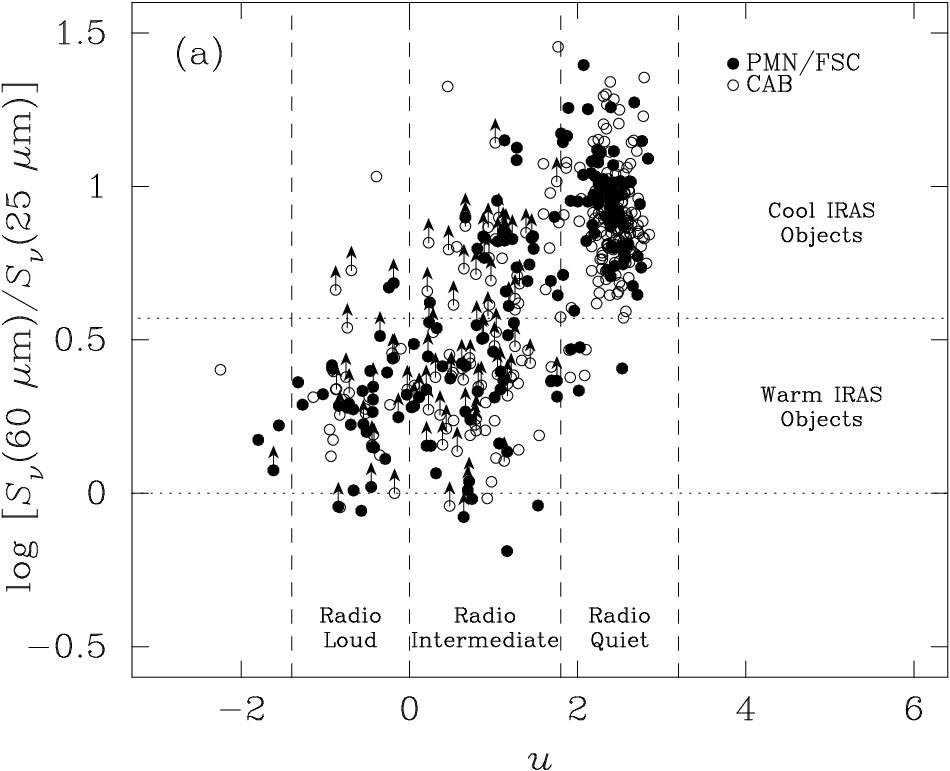}{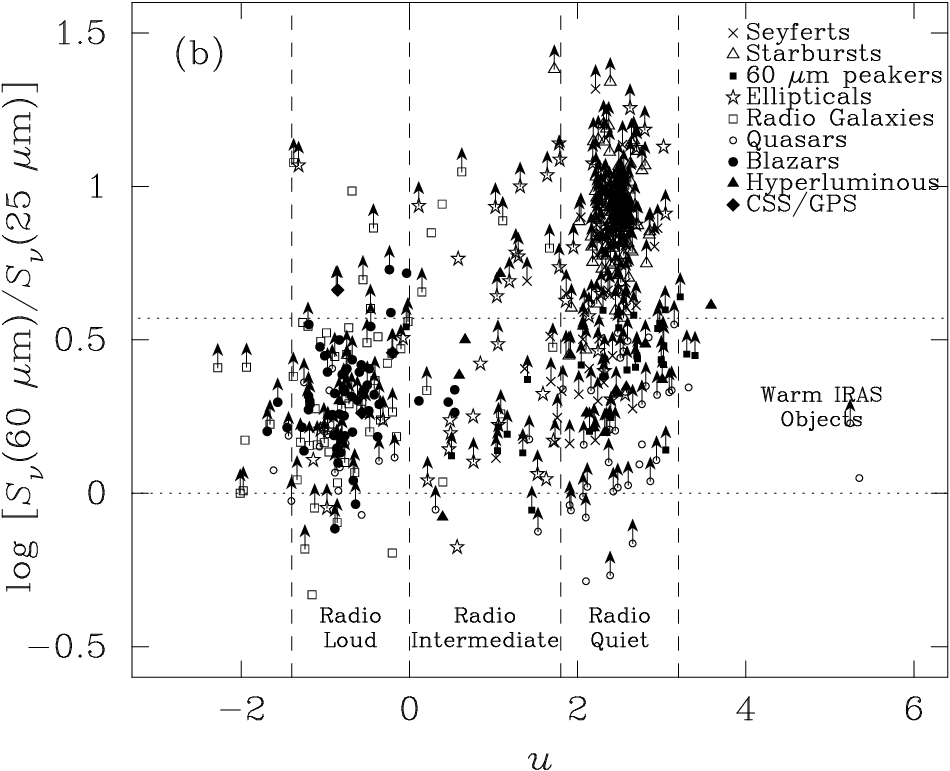}
\caption{(a) FIR color,
$\log~[S_\nu\rm{(60~\micron)}/S_\nu\rm{(25~\micron)}]$, vs radio/FIR
flux ratio, $u = \log~[\Snfir/\Snrad]$, for the full sample.  The
vertical dashed lines define regions occupied by radio-quiet,
radio-intermediate, and radio-loud objects.  The horizontal dotted
lines define regions occupied by ``warm'' and ``cool'' \iras\ galaxies
(de Grijp et al. 1987). (b) FIR color,
$\log~[S_\nu\rm{(60~\micron)}/S_\nu\rm{(25~\micron)}]$, vs radio/FIR
flux ratio, $u$, for the comparison objects (see Table
\ref{tab:com}). The marked regions are the same as for
(a).\label{fig:cols}}
\end{figure}

The \iras\ 25 \micron\ to 60 \micron\ color,
$\log~[S_\nu\rm{(60~\micron)}/S_\nu\rm{(25~\micron)}]$, is a useful
parameter distinguishing starburst galaxies from AGN \citep{gri87,
low88, san88a, gri92}.  Starbursts tend to have cool 25--60 \micron\
color temperatures with
$\log~[S_\nu\rm{(60~\micron)}/S_\nu\rm{(25~\micron)}] > 0.57$ because
of their extended star and gas distributions, while the central
engines in AGN heat the large amounts of nearby dust to warmer
temperatures so $0.0 <
\log~[S_\nu\rm{(60~\micron)}/S_\nu\rm{(25~\micron)}] < 0.57$
\citep{low88}.  It is therefore of interest to compare
$\log~[S_\nu\rm{(60~\micron)}/S_\nu\rm{(25~\micron)}]$ with the radio
excess for each object, which we also use to infer the presence of an
AGN.  Figure \ref{fig:cols}(a) shows this comparison for all objects
in the full sample with 25 \micron\ \iras\ detections.

Figure \ref{fig:cols}(b) shows the colors and radio excesses of the
comparison objects (see Table \ref{tab:com}).  As is well known,
starburst galaxies are radio quiet and have cool far-infrared colors.
In general, the Seyfert galaxies in Figure \ref{fig:cols}(b) have warm
far-infrared colors and small radio excesses.  The classical
radio-loud galaxies and quasars in Figure \ref{fig:cols}(b) have large
radio excesses and almost exclusively have warm far-infrared colors.
The radio-quiet quasars in Figure \ref{fig:cols}(b) have low radio
excesses that are similar to those of the starburst and Seyfert
galaxies, but the far-infrared colors of the radio-quiet quasars are
at least as warm as those of the radio-loud galaxies and quasars.

Our radio-excess sample includes objects with both cool and warm
far-infrared colors, suggesting that the sample may contain objects
powered by both starbursts and AGN.  It will be of interest to
establish whether the radio-intermediate objects with cool
far-infrared colors do indeed have optical spectra indicative of
starbursts, whether these are objects in which the AGN is so heavily
embedded that the emission is optically thick at far-infrared
wavelengths, or whether these objects lack a significant dust mass in
close proximity to the AGN so exhibit only cool dust emission.

There is a tendency in Figure \ref{fig:cols}(a) for objects with
warmer far-infrared colors to also have larger radio excesses (small
$u$ values). However, comparison of Figures \ref{fig:cols}(a) and
\ref{fig:cols}(b) shows that the lower bound of this distribution is
due largely to the absence of radio-quiet quasars in our full
sample. This exclusion is traced to the relatively high radio flux
limits of both the PMN and Greenbank 4.8 GHz surveys: most radio-quiet
quasars fall below the radio flux limits of both our sample and that
of CAB.

Few objects appear in Figure \ref{fig:cols}(a) that are both
radio loud and have cool far-infrared colors. This supports
the association of extreme excess radio emission with an AGN.

\placefigure{fig:cols}
\begin{figure}
\epsscale{0.43}
\plotone{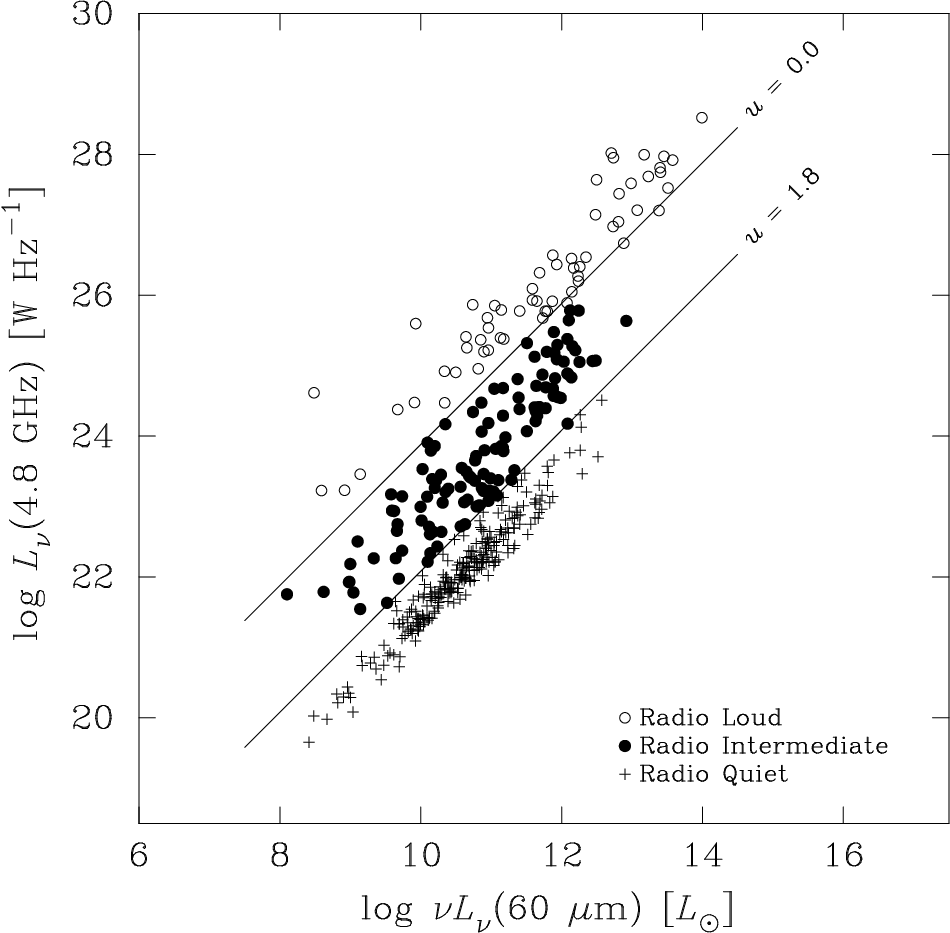}
\caption{Radio/FIR luminosity diagram for the full PMN/FSC and CAB
samples showing radio-loud, radio-intermediate, and radio-quiet
objects with different symbols. Lines corresponding to $u = 1.8$ and
$u = 0.0$ at zero redshift are labelled.\label{fig:udiv}}
\epsscale{1.0}
\end{figure}
\clearpage

\begin{figure}
\epsscale{0.45}
\plotone{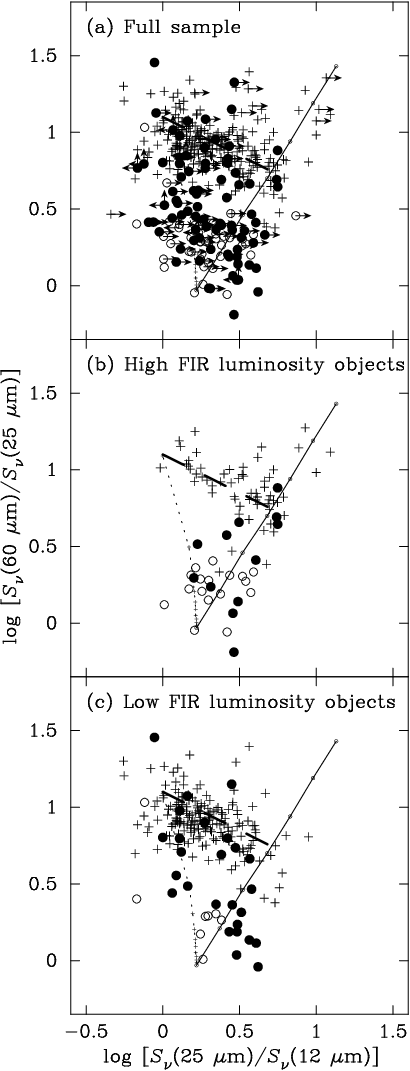} 
\caption{Far-infrared color-color diagram showing the positions of
objects with different radio excesses (as in Fig.\
\ref{fig:udiv}). The top panel shows all PMN/FSC and CAB objects, the
middle panel shows only high FIR luminisity objects with $\nLnfir >
10^{11} \> \Lsun$ and detected at all three FIR wavelengths, and the
bottom panel shows only low FIR luminosity objects with $\nLnfir <
10^{11} \> \Lsun$ and detected at all three FIR wavelengths. The
starburst, mixing, and reddening lines are as shown in Fig.\
\ref{fig:ucols_leg}. The symbols used are as shown in
Fig. \ref{fig:udiv}.\label{fig:ucols}}
\epsscale{1.0}
\end{figure}

\begin{figure}[ht]
\epsscale{0.45}
\plotone{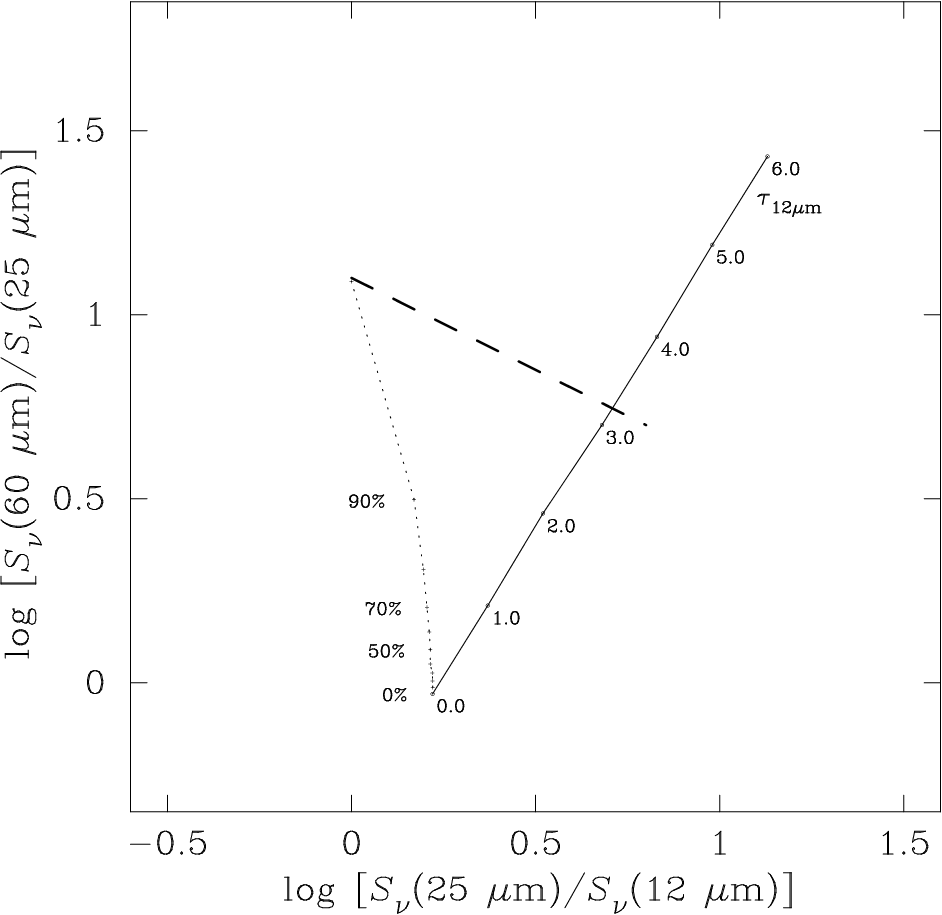}
\caption{Far-infrared color-color diagram showing the mean locus of
starburst galaxies ({\em dashed line}), an extreme mixing line between
a pure Seyfert~1 spectrum and a starburst spectrum ({\em dotted
line}), and the reddening line determined by obscuring a Seyfert~1
spectrum with the 12 \micron\ optical depths that are indicated ({\em
solid line}).\label{fig:ucols_leg}}
\epsscale{1.0}
\end{figure}

Figure \ref{fig:udiv} reproduces the radio/FIR luminosity diagram for
the full PMN/FSC and CAB samples segregated according to $u$ value;
radio-quiet objects ($u \geq 1.8$), radio-intermediate objects ($0.0 <
u < 1.8$), and radio-loud objects ($u \leq 0.0$) are plotted with
different symbols. These objects are shown with the same symbols in
far-infrared two-color diagrams in Figure \ref{fig:ucols}, which
indicates the different color distributions of radio-quiet,
radio-intermediate and radio-loud objects. The distribution of objects
in Figure \ref{fig:ucols} is bounded by a starburst line, a reddening
line, and an extreme mixing line, as explained by \citet{dop98}. These
loci are shown separately for clarity in Figure \ref{fig:ucols_leg}.
The starburst line was derived from the observed colors of starburst
galaxies \citep{dop98}.  Starburst and star-forming galaxies exhibit a
range of $\log~[S_\nu\rm{(25~\micron)}/S_\nu\rm{(12~\micron)}]$ and
$\log~[S_\nu\rm{(60~\micron)}/S_\nu\rm{(25~\micron)}]$ ratios due to a
range of different dust temperatures. Warmer star-forming galaxies
have higher $\log~[S_\nu\rm{(25~\micron)}/S_\nu\rm{(12~\micron)}]$ and
lower $\log~[S_\nu\rm{(60~\micron)}/S_\nu\rm{(25~\micron)}]$ ratios
due to the shift of the blackbody peak to shorter wavelengths
(i.e. towards 25 \micron).  A pure Seyfert~1 nucleus is assumed to
have $\log~[S_\nu\rm{(60~\micron)}/S_\nu\rm{(25~\micron)}] = -0.03$
and $\log~[S_\nu\rm{(25~\micron)}/S_\nu\rm{(12~\micron)}] = 0.22$
\citep{dop98}.  The extreme mixing line is determined by mixing the
unreddened Seyfert~1 spectrum with increasing fractions of ``cool''
starburst spectrum having intrinsic colors of
$\log~[S_\nu\rm{(60~\micron)}/S_\nu\rm{(25~\micron)}] = 0.0$ and
$\log~[S_\nu\rm{(25~\micron)}/S_\nu\rm{(12~\micron)}] = 1.09$. The
reddening law is taken from \citep{dop98}. For each object, the
relative contributions of starburst and AGN components, combined with
the degree of reddening, define its place in the color-color diagram.
\placefigure{fig:udiv}

\placefigure{fig:ucols}

\placefigure{fig:ucols_leg}

The FIR color-color diagram shows differences in the colors of objects
with different degrees of radio excess (Fig. \ref{fig:ucols}).  The
radio-quiet objects in our sample show similar FIR color distributions
to those seen in normal and starburst galaxies \citep{rus93}. The high
FIR luminosity ($> 10^{11} \> \Lsun$) radio-quiet objects extend over
the full range of starburst colors (Fig.\ \ref{fig:ucols}(b)), with a
few objects being distributed along the Seyfert~1 galaxy redddening
line.  The low FIR luminosity ($< 10^{11} \> \Lsun$) radio-quiet
objects (Fig.\ \ref{fig:ucols}(c)) cluster predominantly near one end
of the starburst distribution. This is the region occupied by cool
starburst galaxies in which 11.3 \micron\ PAH emission dominates the
S$_{\nu}$(25 \micron)/S$_{\nu}$(12 \micron) color \citep{dop98}. The
high FIR luminosity objects may contain hotter dust or the destruction
of PAH molecules may be more complete in the high FIR luminosity
objects, possibly due to the greater prominence of radio-quiet
AGN. Radio-loud objects with both high and low FIR luminosities
cluster around the unreddened Seyfert~1 region in Figure
\ref{fig:ucols}. These objects are clearly AGN-dominated with 12
\micron\ optical depths generally below 2.0. In contrast, the
radio-intermediate objects span the color range between the starburst
sequence and the unreddened Seyfert~1 region in this diagram. The high
FIR luminosity radio-intermediate objects are confined to the
Seyfert~1 galaxy reddening line with 12 \micron\ optical depths up to
$\sim$ 4.0 (Fig.\ \ref{fig:ucols}(b)). We infer from this that at high
FIR luminosities radio-intermediate objects are more heavily obsured
than are the radio-loud objects, which have more extreme radio
excesses. The high FIR luminosity radio-intermediate objects occupy
the same region of the FIR color-color diagram occupied by hidden
broad-line region (BLR) objects in which the BLR is revealed by either
spectropolarimetry observations (Heisler et al. 1997) or by
near-infrared emission-line spectroscopy \citep{vei97}. This
similarity suggests that hidden BLRs may be detectable by these
techniques in several of the high FIR luminosity radio-intermediate
objects. The low FIR luminosity radio-intermediate objects extend
along the mixing line between the ``blue'' end of the starburst
sequence and the unreddened Seyfert~1 region. Approximately half of
the low FIR luminosity radio-intermediate objects have FIR colors that
are indistinguishable from those of starburst galaxies. These
differences in the FIR colors of the high and low FIR luminosity
radio-intermediate objects suggest that these objects may have
fundamentally different physical natures.

\subsection{Optical Properties}
\label{sec:dis_opt}

\begin{figure}[ht]
\plottwo{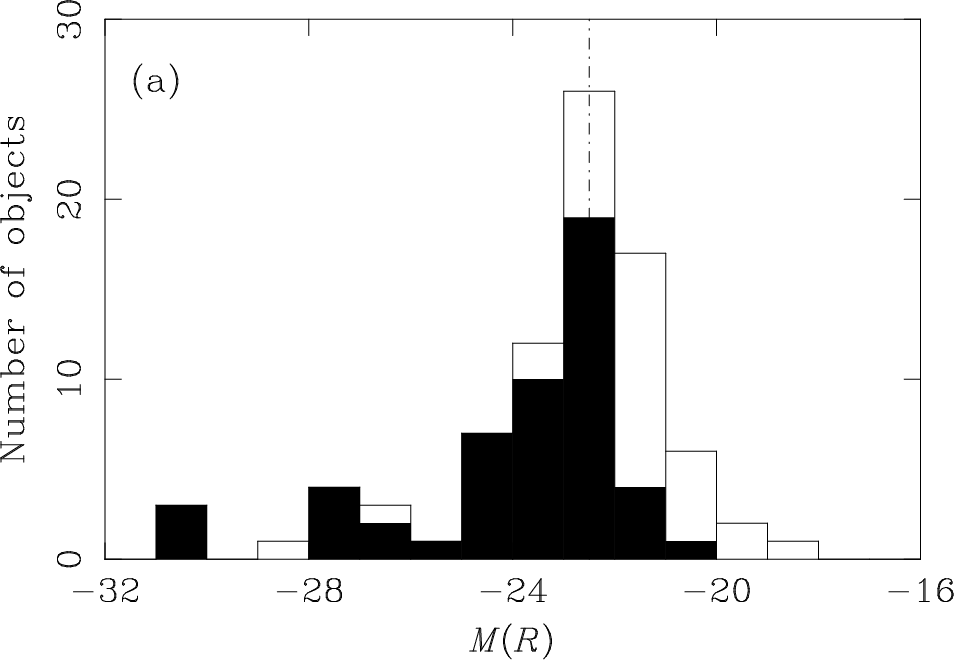}{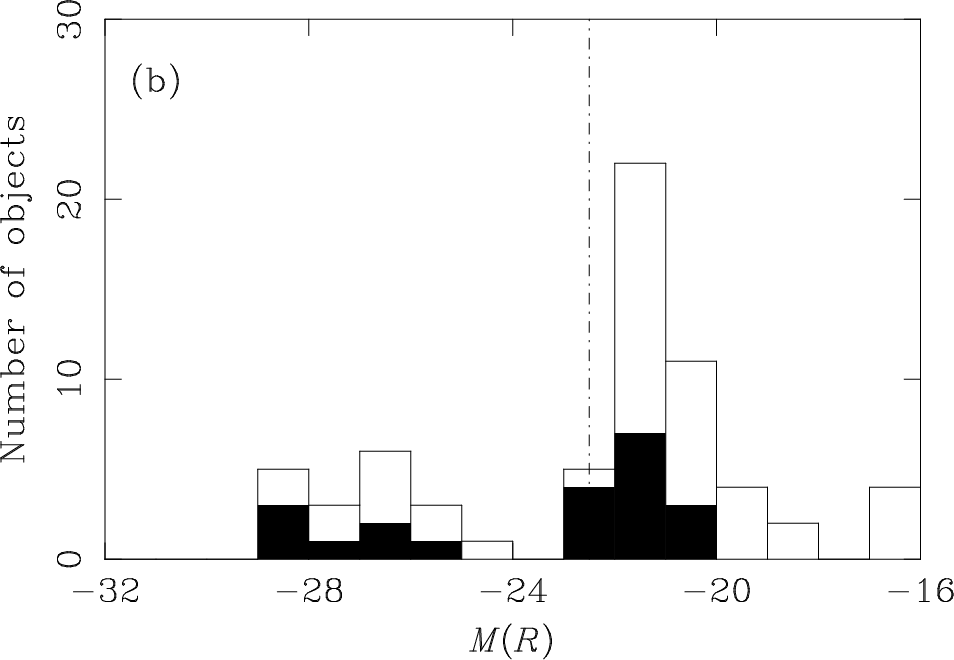}
\caption{Distribution of absolute $R$ magnitude for radio-excess
objects (a) and radio-quiet objects (b) south of declination
+2.5$^\circ$ based on SuperCosmos photographic photometry. Filled
regions indicate high FIR luminosity ($\nLnfir > 10^{11} \> \Lsun$)
objects. Unfilled regions indicate low FIR luminosity ($\nLnfir <
10^{11} \> \Lsun$) objects. The dotted line corresponds to an $L^*$
galaxy with $M(R) = -22.5$ mag \protect\citep{bla01}.\label{fig:mr}}
\end{figure}

SuperCosmos optical photographic
photometry\footnote{http://www-wfau.roe.ac.uk/sss/} has been assembled
for most objects south of declination +2.5$^\circ$ in the PMN/FSC and
CAB samples. The $R$ magnitudes are accurate to only $\sim$ 0.5 mag
\citep{ham01}. However, this is sufficient to derive indicative total
absolute magnitudes, $M(R)$. The distribution of $M(R)$ is shown in
Fig.\ \ref{fig:mr}(a) for the radio-excess objects with SuperCosmos
$R$ magnitudes and known redshifts. The optical counterparts to these
radio-excess objects are all high luminosity galaxies with $M(R)$
comparable to or brighter than that of an $L^*$ galaxy (indicated by a
dotted line in Fig.\ \ref{fig:mr}). This is consistent with other
evidence that radio activity is associated with massive host galaxies
\citep{dun01}.  There is a suggestion in Fig.\ \ref{fig:mr}(a) that
the high FIR luminosity objects also have higher optical
luminosities. Calibration uncertainties in the photographic photometry
are such that CCD photometry will be required to verify this result.
\placefigure{fig:mr}

The magnitude distribution for the optical counterparts of the
radio-quiet objects is shown in Figure \ref{fig:mr}(b). The peak of
the radio-quiet objects is about 1.5 magnitudes fainter than that of
the radio-excess objects. The distribution of radio-quiet objects also
extends to fainter optical magnitudes. This further supports the
suggestion that the radio activity is associated with massive,
luminous galaxies, although more accurate photometry is required to
confirm this result.

\section{Conclusions}
\label{sec:con}

The \iras\ FSC 60 \micron\ data have been cross-correlated with the
PMN 5 GHz catalog to identify objects that emit significantly at both
radio and FIR wavelengths. A significant population of radio-excess
objects has been found that lies above the radio-quiet radio/FIR
correlation. We find that the fraction of radio-intermediate objects
in our sample is much larger than has been found by previous authors,
especially at high FIR luminosities. This is because we have probed to
lower \iras\ flux densities. Future surveys by SIRTF are expected to
discover more of these objects. The radio-excess objects appear to be
uniformly distributed in the radio/FIR luminosity diagram between the
radio-quiet sequence and the extreme radio-loud limit. There is no
evidence for a separate radio-loud sequence. The apparent division of
AGN into distinct radio-quiet and radio-loud classes appears to be due
to the high flux density limits of previous radio surveys. This also
affects our sample in that objects with small radio excesses and faint
60 \micron\ flux densities are excluded because they fall below the
PMN flux limit.

Many of the radio-excess objects have optically thin radio emission on
scales $\le$ 15~kpc, likely to be smaller than the host galaxy. These
properties are similar to CSS/GPS sources, but the radio-excess
objects have lower radio powers that are closer on average to those of
Seyfert radio sources than CSS/GPS sources.  Higher resolution radio
data are required to determine the morphologies and sizes of the
radio-excess sources. More spectral energy distribution data are also
needed to determine if the spectra of the radio-excess objects turn
over at GHz frequencies. This will allow us to more confidently test
whether the radio-excess objects are low power CSS/GPS sources, with
the same physical processes driving the radio emission as in these
objects.

Differences between the FIR colors of the radio-excess objects with
low and high FIR luminosities suggest they may be different types of
objects. Some low far-infrared luminosity radio-excess objects appear
to derive a dominant fraction of their far-infrared emission from star
formation, despite the dominance of the AGN at radio
wavelengths. Amongst the high FIR luminosity objects, we find that
those with higher radio excesses tend to have FIR colors typical of
unobscured Seyfert 1 galaxies, and objects with smaller radio excesses
have colors indicating higher obscuration. If the radio power, and
hence the radio excess, increases with age this may suggest that
younger sources are more obscured than older sources.  Our
radio-excess sample contains high FIR luminosity objects over the full
range of radio excesses.  Well-studied interacting Seyfert galaxies
with compact radio emission are present among the radio-quiet and
radio-intermediate members of the sample.  At high FIR luminosities,
these are caused by the mergers of large dusty spiral galaxies, such
as the Superantennae and Mrk 463.  They contrast with the powerful
isolated radio galaxies, such as 3C~195, which have large radio
excesses and may be associated with the infall of a smaller dusty
companion into an elliptical galaxy. These powerful radio sources may
also be in a more advanced merger state than the less luminous radio
sources that still exhibit two separate optical nuclei. This suggests
that the merging activity may be associated with the development of
the radio source.  It is clearly of interest to obtain higher quality
optical images of the radio-excess galaxies. This would reveal the
true morphologies of the host galaxies, and allow us to determine what
fraction of the radio-excess objects are interacting and whether they
are in an early or late merger stage.

The properties of many of the radio-excess objects are consistent with
an evolutionary picture in which young radio sources are emerging from
a dusty nucleus, and increasing in radio power as they evolve. Objects
with \nLnfir $\gtrsim 10^{11} \> \Lsun$ are classified as luminous and
ultra-luminous infrared galaxies.  Such high FIR luminosities can be
powered by merging activity between galaxies.  It is known from
observations \citep{san86, san88c} and numerical simulations
\citep{bar96} that tidal interactions between galaxies can funnel gas
towards the nucleus, which may trigger bursts of star
formation. Interactions and mergers have also been associated with the
fuelling of AGN activity and with the generation of powerful radio
emission \citep{hec86}, although strong tidal interaction does not
necessarily produce a radio-loud AGN \citep{smi93}. If the
radio-excess objects are low-power CSS/GPS sources, they are likely to
be young radio sources, triggered by tidal interactions that are also
responsible for the high FIR luminosities. In this case, even if the
radio-excess sources were initially in a ``GPS phase'', increasing in
radio power as they evolved, they will eventually fade as the radio
lobes expand.  It is possible that they may fade to radio-quiet
objects or weak FR~I's.

Alternatively, it is possible that the radio sources may be confined
by the high gas density in the nuclear region, FIR-luminous galaxies
have higher gas masses than observed in CSS/GPS sources, which may be
sufficient to confine the radio jets.  The weaker radio powers of the
radio-excess objects may be due to intrinsically weaker radio jets
which would be easier to confine than powerful CSS/GPS sources.
Several observational tests may be made to determine whether the FIR
luminous radio-excess objects are likely to be young or
confined. Observations to detect \ion{H}{1} absorption or CO emission
would provide estimates of the gas masses in the hosts, which could
indicate whether confinement is possible.  A study of the radio
spectral energy distributions of the objects may enable an estimate of
the age of the emitting electrons via synchrotron cooling.
Measurement of the \oiii\ luminosities will allow an estimate of the
jet energy fluxes, which may be compared with those of Seyfert radio
sources and CSS/GPS sources.

Optical spectroscopy will also be useful to confirm the presence of an
AGN and to assess how dominant the AGN is in the optical
regime. Furthermore, if these really are CSS/GPS sources, it may be
useful to distinguish the quasars from the narrow-line objects, as it
has been suggested that these two classes of CSS/GPS sources are
physically different types of object that share only a similar
spectral shape \citep{sne00a}. The FIR colors of the high luminosity
objects suggest large obscurations even at FIR wavelengths; optical
spectra may be used to determine the optical extinction which should
be high. The high FIR luminosities of these objects may indicate
merging activity, which would be expected to be accompanied by
starburst activity. Two of the well-studied CSS sources in the sample,
3C~459 and 3C~48, show optical absorption spectra and blue continua
that are indicatative of a young stellar population.  This lends
support to the idea that the radio sources are young, rather than
pressure-confined. Similarly, optical spectra of the radio-excess
objects may be used to determine if there has been recent star
formation and estimate the age of the dominant stellar population.

\acknowledgements

We thank Lisa Kewley for help at an early stage of this research and
Geoff Bicknell for useful discussions. The assistance of the duty
astronomers at the Australia Telescope Compact Array is much
appreciated. Drake is the holder of an Australian Postgraduate
Research Award and a Duffield Scholarship.  MD wishes to acknowledge
the support of the Australian National University and the Australian
Research Council (ARC) through his ARC Australian Federation
Fellowship, and also under ARC Discovery project DP0208445.  The
Australia Telescope is funded by the Commonwealth of Australia for
operation as a National Facility managed by CSIRO.  This research has
made use of the NASA/IPAC Extragalactic Database (NED) which is
operated by the Jet Propulsion Laboratory, California Institute of
Technology, under contract with the National Aeronautics and Space
Administration.  The Digitized Sky Surveys were produced at the Space
Telescope Science Institute under U.S. Government grant NAG
W-2166. The images of these surveys are based on photographic data
obtained using the Oschin Schmidt Telescope on Palomar Mountain and
the UK Schmidt Telescope. The plates were processed into the present
compressed digital form with the permission of these institutions.
IRAF is distributed by the National Optical Astronomy Observatories,
which are operated by the Association of Universities for Research in
Astronomy, Inc., under cooperative agreement with the National Science
Foundation.

\protect\singlespace

\end{document}